\documentclass[twocolumn]{aastex63}
\usepackage[utf8]{inputenc}
\usepackage{graphicx}
\usepackage{amsmath}
\usepackage{amssymb}
\usepackage{xcolor}
\usepackage{multirow}
\usepackage{pifont}
\newcommand{\xmark}{\ding{55}}

\shorttitle{Binary-Stripped CCSNe}
\shortauthors{Vartanyan et al.}

\begin{document}

\title{Binary-Stripped Stars as Core-Collapse Supernovae Progenitors}

\correspondingauthor{David Vartanyan}
\email{dvartany@berkeley.edu}

\author[0000-0003-1938-9282]{David Vartanyan}\affiliation{Department of Physics and Astronomy, University of California, Berkeley, CA  94720, USA}
\author[0000-0003-1009-5691]{Eva Laplace}
\affiliation{Anton Pannekoek Institute of Astronomy and GRAPPA, Science Park 904, University of Amsterdam, 1098XH Amsterdam, The Netherlands}
\author[0000-0002-6718-9472]{Mathieu Renzo}
\affiliation{ Department of Physics, Columbia University, New York, NY 10027, USA}
\affiliation{Center for Computational Astrophysics, Flatiron Institute, New York, NY 10010, USA}
\author{Ylva G\"otberg}
\affiliation{The Observatories of the Carnegie Institution for Science, 813 Santa Barbara Street, Pasadena, CA 91101, USA}
\author{Adam Burrows}
\affiliation{Department of Astrophysical Sciences, 4 Ivy Lane, Princeton University, Princeton, NJ 08540, USA}
\author{Selma E. de Mink}
\affiliation{ Max Planck Institute for Astrophysics, Karl-Schwarzschild-Str. 1, 85748 Garching, Germany}
\affiliation{Anton Pannekoek Institute of Astronomy and GRAPPA, Science Park 904, University of Amsterdam, 1098XH Amsterdam, The Netherlands}
\affiliation{Harvard-Smithsonian Center for Astrophysics,
60 Garden St., Cambridge, MA 02138, USA}

\date{January 2020}

\begin{abstract}
Most massive stars experience binary interactions in their lifetimes that can alter both the surface and core structure of the stripped star with significant effects on their ultimate fate as core-collapse supernovae. However, core-collapse supernovae simulations to date have focused almost exclusively on the evolution of single stars. We present a systematic simulation study of single and binary-stripped stars with the same initial mass as candidates for core-collapse supernovae (11 $-$ 21 M$_{\odot}$). Generally, we find that binary-stripped stars core tend to be less compact, with a more prominent, deeper silicon/oxygen interface, and explode preferentially to the corresponding single stars of the same initial mass. Such a dichotomy of behavior between these two modes of  evolution  would have important implications for supernovae statistics, including the final neutron star masses, explosion energies, and nucleosynthetic yields. Binary-stripped remnants are also well poised to populate the possible mass gap between the heaviest neutron stars and the lightest black holes. Our work presents an improvement along two fronts, as we self-consistently account for the pre-collapse stellar evolution and the subsequent explosion outcome. Even so, our results emphasize the need for more detailed stellar evolutionary models to capture the sensitive nature of explosion outcome. 
\end{abstract}  
\keywords{
stars - supernovae - general; binaries - close}

\section{Introduction}

Studies on the core-collapse supernovae (CCSNe) explosion mechanism have focused almost exclusively on single-star
progenitors \citep{roberts2016,radice2017b,summa2018,ott2018_rel, vartanyan2018a,vartanyan2018b,burrows2018,burrows_2019,burrows_2019b,2021Natur.589...29B,oconnor_couch2018b,nagakura2019,glas2019,kuroda2020}. However, the vast majority of stars massive enough to
reach core-collapse are members of multiple systems \citep[e.g.,][]{mason:09, almeida:17}. \citet{sana:12} conclude that the majority of massive stars should interact with a close companion during their lifetime. Binary evolution is commonly required to explain the high intrinsic rate of hydrogen poor (type IIb, Ib, Ic) CCSNe \citep{Podsiadlowski+1992,Eldridge+2008,2010ApJ...725..940Y,li_nearby_2011,Claeys+2011,2017ApJ...842..125Z,
Shivver2019, Sravan+2019} and their ejecta mass distributions peaked at low values of about 2 M$_{\odot}$ \citep[e.g.,][]{lyman_bolometric_2016,taddia_carnegie_2018}. It is also required to explain the properties of several SN progenitors observed through direct imaging, including the triple-ring structure of SN1987a \citep[e.g.,][]{eldridge_death_2013,eldridge_binary_2017,2021arXiv210209686U}. Moreover, direct imaging provides evidence for surviving companions after CCSNe \citep[e.g.,][]{maund_massive_2004,fox_uncovering_2014,ryder_ultraviolet_2018}. Binary population synthesis studies have recently shown binary interaction should also affect $\sim$\,$50\%$ of hydrogen-rich (type II) SNe \citep[e.g.,][]{eldridge:18,zapartas_diverse_2019,2021A&A...645A...6Z}.

Thus, we expect that a large fraction of CCSN progenitors
experiences binary interactions, which might change the
core structure \citep{1989A&A...210...93L,1993ApJ...411..823W,laplace21} and thus ``explodability'' of stars of a given initial mass. This is buttressed by a growing anthology of observed stripped SNe (\citealt{Shivver2019}) with interesting implications for the formation of double compact objects (\citealt{de2018,2018A&A...609A.136T,2019MNRAS.485.1559P}) and rapidly evolving transients (\citealt{2018MNRAS.481..894P}).

CCSNe explosion simulations of progenitors computed accounting self-consistently
for binary interactions are as-yet rare, as are detailed stellar evolution models of binary stars computed to core-collapse. The preponderance of studies to date focus rather on the structure of naked cores (naked He cores, \citealt{woosley2019}; and naked CO cores, \citealt{patton2020}, but see \citealt{tauris:15}), and prompt explosion through parametrized or prescriptive means \citep{woosley2019,2021A&A...645A...6Z}. Our intent here is to provide a more detailed approach encompassing both binary progenitor formation and subsequent explosion outcome.

The work of
\cite{mueller:18, mueller:19} considered the impact of binary evolution
modeling on the explodability of the donor star with 3D explosion
simulations, although with several approximations to the neutrino radiation transport and binary evolution modeling. They focused on the explosion of ultra-stripped stars from
\cite{tauris:15}, motivated by the relevance of this channel for gravitational-wave progenitors. The binary evolution models from \cite{tauris:15} start from
initial conditions with a compact object (NS) orbited by a naked He star, i.e.
starting the evolution from after the first mass transfer or common envelope
phase in a binary. The binary evolution leads to a second mass transfer phase
(case BB RLOF, Roche-lobe overflow) which further reduces the mass of the star, making it
ultra-stripped. They found that these ultra-stripped stars blow up with weak
explosion energies ($E\simeq10^{50}\,\mathrm{erg}$) and with prompt explosions
resulting in small SN natal kicks.

Our study, using the progenitors developed in \cite{laplace21}, is complementary to the work of \cite{mueller:18, mueller:19} and
\cite{tauris:15}, since we focus instead on the first RLOF phase and what impact
it can have on the explodability of stripped (but not ultra-stripped) CCSN progenitors. The most common kind of
binary interaction is mass transfer with a post main sequence donor (so-called
case B RLOF, \citealt{2008AIPC..990..230D,Klencki+2020}), which is predicted to leave only
a very thin layer of H-rich material on the donor star at the end of the mass
transfer phase at solar metallicity \citep{2017ApJ...840...10Y,gotberg:17,Gilkis+2019}. At solar
metallicity, such a layer is likely removed by stellar winds in the post-mass transfer, pre-collapse
evolution, leaving an exposed He core which will likely be the progenitor of a Ib SN explosion, or, if the helium-rich layers are removed, of a Ic. Therefore, albeit with numerous
caveats affecting any stellar evolution simulations, our self-consistent binary
models and supernovae simulations are the first used to explore the impact of the most common binary evolution channel, case B mass transfer, on the explodability of the donor stars
core with the added advantage of using a more-detailed radiation transport scheme to simulate the resulting supernovae. 

Here, we report on the first comprehensive, self-consistent simulation study comparing CCSNe outcomes for a suite of 11 single-star and binary-stripped progenitors with the same initial mass \citep{laplace21}. This allows for the first time a systematic study of the viability of stripped stars as CCSNe progenitors and explores the role of massive star evolutionary history on CCSNe outcome. We find preferential explosion of binary-stripped progenitors over single-star progenitors of the same ZAMS mass. After describing the salient points of our pre-explosion and explosion modeling in Sec.\,\ref{sec:methods}, we first compare a representative pair of models with the same initial mass in Sec.\,\ref{sec:properties}. We find that overall donor stars in binaries are easier to explode, and among pairs of the same initial mass, lower compactness corresponds to easier explodability, but highlight key exceptions to this general trend and present explosion diagnostics with observational ramifications. We present our conclusions in Sec.\,\ref{sec:conc}.


\section{Methods}
\label{sec:methods}
\subsection{MESA}

The stellar evolution models are presented and described in detail in \cite{laplace21}. The models were computed with the open-source 1D stellar evolution code MESA \citep[version 10398,][]{paxton_modules_2011,paxton_modules_2013,paxton_modules_2015,paxton_modules_2018,paxton_modules_2019-1} at solar metallicity \citep[$Z=0.0142$,][]{asplund_chemical_2009}. Until core oxygen depletion (central oxygen mass fraction lower than $10^{-4}$), we used a nuclear network comprising 21 isotopes that produces sufficiently accurate reaction rates in the cores of massive stars from core hydrogen burning until the end of core oxygen burning \citep[\texttt{approx21},][]{timmes1999,timmes2000,paxton_modules_2011}. Because of the sensitivity of the further evolution to nuclear burning and especially electron captures, we used a 128-isotope network after oxygen depletion \citep{farmer_variations_2016} and stop our models when the iron-core infall velocity reached 1000 km s$^{-1}$.

Our grid consists of two sets of 11 models with the same initial masses $M_1$ ranging from 11 to 21 M$_{\odot}$. The first set follows the evolution of single stars. For the second set, we considered the most common binary interaction, that is, a stable mass transfer phase after the end of the donor's main sequence \citep[case B mass transfer, e.g.,][]{kippenhahn_entwicklung_1967}. We examined binaries with initial periods of  25--35 d and a point-mass secondary of mass $M_2 = 0.8 M_1$, and assumed fully conservative mass transfer.
All stellar models and MESA inlists are publicly available upon publication (see \citealt{laplace21}).

\begin{figure*}
    \centering
  \includegraphics[width=0.44\textwidth]{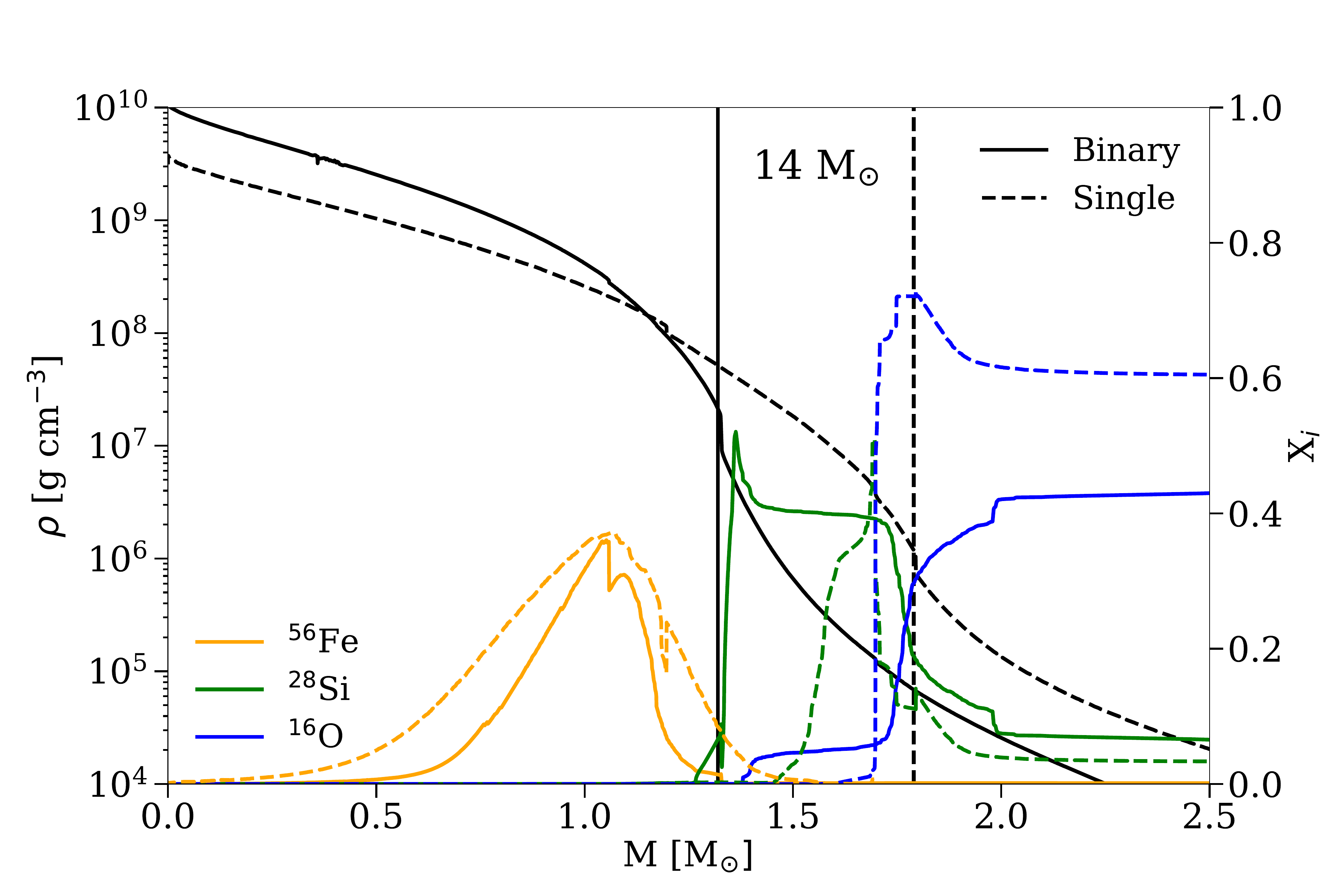}
 \hspace*{0.1cm}
	\includegraphics[width=0.54\textwidth,height=0.224\textheight]{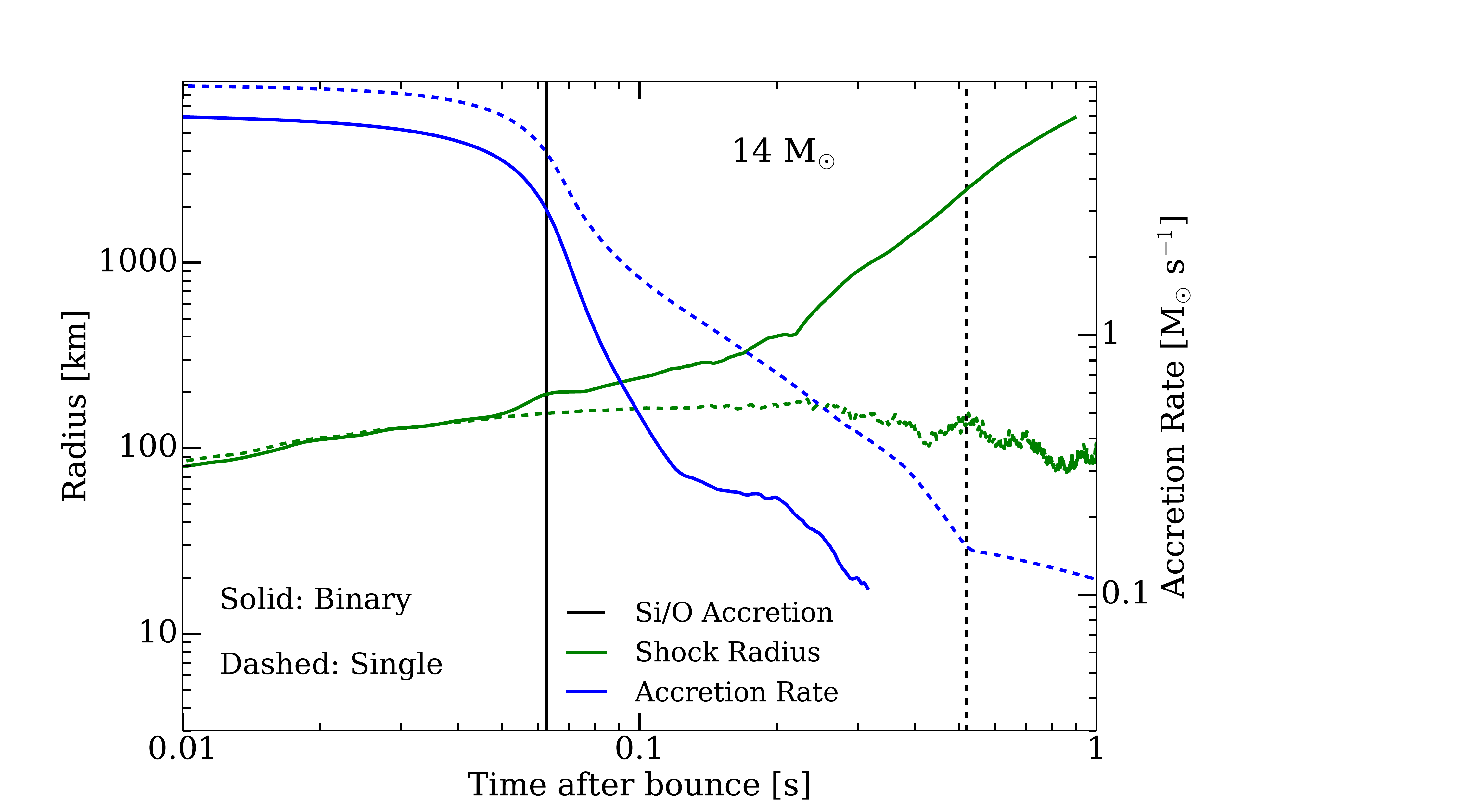}
    \hspace*{-1.2cm}
    \includegraphics[width=0.49\textwidth]{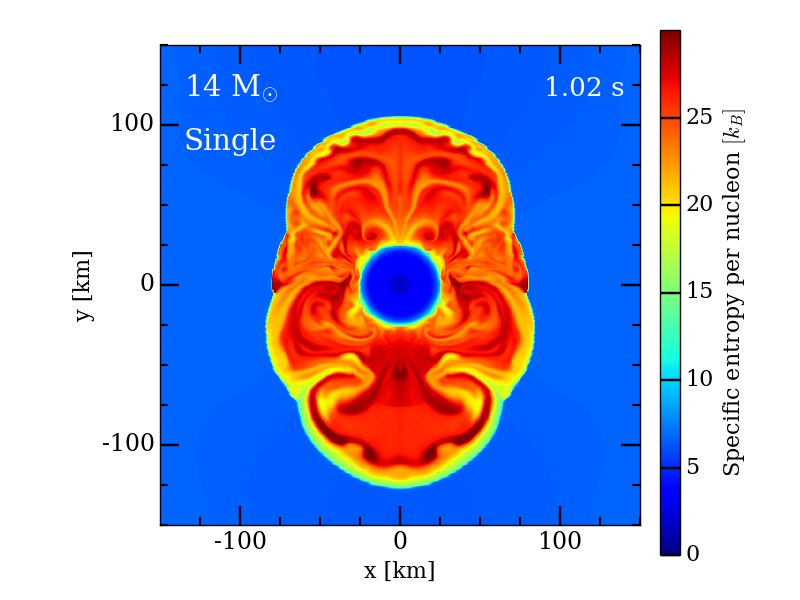} 
    \hspace{-.8cm}
 \includegraphics[width=0.49\textwidth]{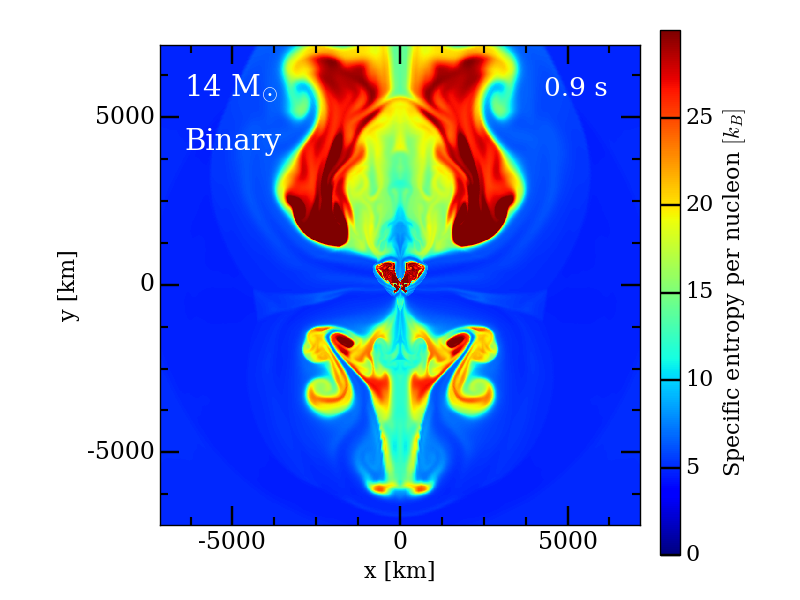}  
    
	\caption{\textbf{Top Left}: Initial density profiles (black) and the mass fraction of Fe-56 (red), Si-28 (green), and O-16 (blue) distributions in the interior 2.5 M$_{\odot}$ for the binary-stripped (solid) and single (dashed) 14-M$_{\odot}$ model. The vertical solid and dashed lines indicate the compositional interface, respectively.  \textbf{Top Right}: Mean shock radii (km, green) and the accretion rate at 200 km, just exterior to the shock, (M$_{\odot}$ s$^{-1}$, right y-axis) for the 14-M${_\odot}$ progenitor, a typical binary/single pair as a function of time after bounce (s). The vertical lines illustrate when the Si/O interface is accreted onto the expanding shock. For both cases, the accretion of the Si/O interface corresponds to the onset of turbulence in the expanding shock front, illustrated by variations in the mean shock radii. The concurrent accretion rate plummets as a result of the drop in density and pressure outside the interface. The shock begins to expand rapidly for the binary-stripped model, resulting in an explosion. However, the Si/O interface for the single star is further out and less pronounced, and proving insufficient to revive the already stalled shock. Nevertheless, even the single-star model shows a bump in the shock radii shortly after accretion of the Si/O interface. \textbf{Bottom}: We show the two-dimensional entropy profiles of the  14-M$_{\odot}$ model for the non-exploding single star on the left and the exploding binary-stripped star on the right. Note the vastly different scales plotted.}
	\label{fig:typical}
\end{figure*}

\subsection{{FORNAX}}
To simulate the collapse, core-bounce, and initial shock propagation in the first seconds, we use {F\sc{ornax}} \citep{skinner2019}. {F\sc{ornax}} is a multi-dimensional, multi-group radiation hydrodynamics code originally
constructed to study core-collapse supernovae. The models referenced herein were evolved in 2D axisymmetry on a spherical grid extending to 20,000 km and resolved with 678 radial cells and 256 angular cells. The angular grid resolution varies smoothly from 0.64$^{\circ}$ along the equator to 0.94$^{\circ}$ along the poles. Following \cite{marek2006} we used a monopole approximation for relativistic gravity and employed the SFHo equation of state (\citealt{2013ApJ...774...17S}), which is consistent with all currently known constraints (\citealt{2017ApJ...848..105T}) on the nuclear equation of state. Our intention in this study is to identify trends in a large set of models, and hence simulating in 2D axisymmetry is favorable. Although 3D simulations still are required, earlier works show similarities in explosion outcome and diagnostics between 2D and 3D simulations (e.g, \citealt{vartanyan2018a,burrows_2019b,2021Natur.589...29B}), and such a broad suite of 2D simulations as presented here sets the groundwork for more selective 3D simulations in the future. 

We solve for radiation transfer using the M1 moment closure scheme for the second and 
third moments of the radiation fields (\citealt{2011JQSRT.112.1323V}) and follow 
three species of neutrinos: electron-type ($\nu_{e}$), anti-electron-type ($\bar{\nu}_{e}$), 
and ``$\nu_{\mu}$"-type ($\nu_{\mu}$, $\bar{\nu}_{\mu}$, $\nu_{\tau}$, and $\bar{\nu}_{\tau}$ neutrino species collectively). We use 12 energy groups spaced logarithmically between 1 and 300 MeV for the electron neutrinos and to 100 MeV for the anti-electron- and ``$\nu_{\mu}$"-neutrinos. The M1 solver avoids simplification to the neutrino transport, such as the fast-multigroup transport scheme with the ray-by-ray approximation, which introduces numerical artifacts into explosion outcome.


\section{Results}
\label{sec:properties}

We present a study of 11 pairs of progenitors, for a total of 22 models, corresponding to progenitors, spanning 11 $-$ 21 M$_{\odot}$ ZAMS mass and following single and binary-stripped post-main sequence evolution. 
We find that all but four of the 22 models explode, where we identify explosion as the shock reaching runaway expansion and failed explosion as the shock stalling. Models explode between within 100 $-$ 800 ms post-bounce. The 13- and 14-M$_{\odot}$ single-star models fail to explode, as do the 17- and 21-M$_{\odot}$ binary-stripped progenitors. 

We compare here a typical single star with a binary-stripped star  with the same ZAMS mass and identify the explosion trends. We then highlight the exceptions below. We plot in Fig.\,\ref{fig:typical} the evolution of the 14-M$_{\odot}$ progenitor pair as a case study of the differences between binary-stripped and single star evolution. The shock radii and density profiles of all progenitors studied are summarized in the \hyperref[app]{Appendix} in Fig.\,\ref{fig:shock} and Fig.\,\ref{fig:rho_M_rshock}, respectively. In the top left panel of Fig.\,\ref{fig:typical}, we illustrate the chemical composition of the interior 2.5 M$_{\odot}$ for the binary-stripped/single star we pair of the same initial mass. We identify a composition interface $-$ which often corresponds to the location of the silicon/oxygen (Si/O) transition $-$ illustrated by a sharp drop in the density, as one metric of explodability (see also \citealt{vartanyan2018a}).\footnote{Categorically, this transition has been associate with a silicon/oxygen boundary (see also \citealt{fryer1999, ott2018_rel}). However, we found that the density drop corresponding to a compositional interface, especially if fragmented, could also correspond to iron/silicon or oxygen/neon/magnesium boundaries. We refer to the Si/O interface, and more broadly, the silicon-group and oxygen-group compositional boundary, interchangeably. See See \hyperref[app]{Appendix} and Fig.\,\ref{fig:rho_M_rshock} for more detail.} The Si/O interface is more pronounced for the binary-stripped progenitor, where the density drops by a factor of $\sim$2.5 over an annulus of $\sim$0.005 M$_{\odot}$, than the single star progenitor, which shows a density drop of only $\sim$1.6 at the interface. Additionally, the interface is located deeper in, at $\sim$1.3 M$_{\odot}$ for the binary-stripped, than for the single star, at $\sim$1.8 M$_{\odot}$.
 
We illustrate the shock radii and highlight the time of Si/O interface accretion and corresponding accretion rate in the top right panel of Fig.\,\ref{fig:typical}. The binary-stripped star has a sharper Si/O interface located deeper within the stellar progenitor than the single-star model (by $\sim$0.5 M$_{\odot}$ for this progenitor mass), and hence the interface is accreted earlier by the expanding shockwave. For the binary-stripped model, the shock intersects the Si/O interface within the first 100 ms. The accretion rate plummets, the ram pressure exterior to the shock drops, and the shock is revived. For the single-star analog, the Si/O interface is less sharp and located further out. It is accreted at $\sim$500 ms post-bounce, 400 ms after the the binary-stripped model. The drop in accretion rate, and hence ram pressure, is noticeably smaller. The single-star 14-M$_{\odot}$ progenitor does show a small bump in the shock radii at the time of Si/O accretion and short-timescale variations as turbulence develops, but the accretion of the Si/O interface is insufficient to revive the stalled shock. Note that in both models, the shock radii are very similar until $\sim$70 ms post-bounce. Stellar collapse and shock revival proceeds quasi-spherically until this point, when turbulence develops around the shock front (\citealt{2013ApJ...778L...7C}). The evolution paths diverge in part due to the different compositional interfaces. We find that the binary-stripped models studied here typically have a sharper interface located deeper in the stellar interior and are also less compact than their single-star counterparts, and, as a result, are more explodable (see Table\,\ref{sn_tab}). 

In the bottom panels of Fig\,\ref{fig:typical}, we plot entropy profiles for the chosen 14-M$_{\odot}$ pair. The single-star shock stalls interior to 100 km roughly one second after bounce, whereas it has reached almost 10,000 km for the exploding binary-stripped star. We see typical entropies as high as 30 k$_\mathrm{B}$ per nucleon, with the explosion occupying a volume-filling fraction of $\sim$20\% at late times. Note the large dipolar asymmetry seen in the exploding model in bottom right panel. Although we did see similar `wasp's waist' morphologies of explosion in various 3D simulations \citep{vartanyan2018b,burrows_2019b}, 3D simulations seem to be less asymmetric than 2D equivalents. 

\begin{table*}
\caption{Explosion Properties}
\centering
\hspace{-2cm}
\begin{tabular}{*{9}{p{11mm}}}
    \hline\hline
  Model        & \multicolumn{2}{c}{Explosion?} & \multicolumn{2}{c}{Compactness}  & \multicolumn{2}{c}{PNS Mass (M$_{\odot}$)} & \multicolumn{2}{c}{He Mass (M$_{\odot}$)} \\
            (M$_{\odot}$)    & Binary &  Single & Binary & Single & Binary & Single & Binary & Single  \\\hline
11.0    &    \checkmark\checkmark     &    \checkmark    & 0.140  & 0.215  & 1.47 & 1.48 & 3.16 & 3.78 \\
12.1   &  \checkmark       &     \checkmark\checkmark  & 0.246 & 0.110 & 1.52 & 1.33 & 3.93 & 4.78 \\
13.0   &  \checkmark\checkmark    &     \xmark       & 0.213 & 0.364 & 1.50 & 1.83 & 4.29 & 5.26 \\
14.0   & \checkmark\checkmark       &   \xmark   & 0.200 & 0.597 & 1.45 & 1.86 & 4.53 & 5.62 \\
14.6  & \checkmark\checkmark & \checkmark     & 0.280 & 0.656 & 1.56 & 1.86 & 4.66 & 5.80  \\
15.0 & \checkmark\checkmark & \checkmark & 0.285 & 0.478 & 1.53 & 1.75 & 5.03 & 6.32 \\
16.0 & \checkmark\checkmark & \checkmark & 0.274 & 0.647 & 1.52 & 1.89 & 5.36 & 6.85\\
17.0 & \xmark & \checkmark\checkmark  & 0.552 & 0.740 & 1.85 & 1.99 & 5.66 & 7.30 \\
18.0 & \checkmark & \checkmark\checkmark & 0.569 &  0.756  & 1.79 & 2.01 & 6.01 & 7.41 \\
20.0   & \checkmark\checkmark        & \checkmark  & 0.678 & 0.723  & 1.92 & 2.16 & 6.34 & 7.57    \\
21.0   &     \xmark   &      \checkmark\checkmark & 0.730 &  0.363 & 2.16 & 1.61 & 6.67 & 8.19 \\
\hline
  \end{tabular}
  \\\begin{flushleft}Table of our 2D simulation results: models with a checkmark explode, and models with an \xmark\, do not explode. Model with two checkmarks explode first compared to their single/binary counterpart, if the latter explodes at all. We also show the compactness (at 1.75 M$_{\odot}$) and the final PNS mass (in M$_{\odot}$) for the binary-stripped and single star pairs, as well as the helium core masses at the onset of core-collapse. Note that these PNS masses are lower limits $-$ many of the models continue to accrete (and explode) at the end of our simulations. \end{flushleft}
  \label{sn_tab}
\end{table*}

\subsection{Compactness}

We find that the compactness parameter provides a viable relative metric of explodability for a given binary/single-star pair of the same ZAMS mass. The compactness parameter characterizes the core structure and is defined as \citep{2011ApJ...730...70O}:

\begin{equation}
\xi_M= \frac{M/M_{\odot}}{R(M)/1000\, \mathrm{km}}\,\,
\end{equation}

where the subscript $M$ denotes the mass where compactness is evaluated. For our purposes, we evaluate at $M$ = 1.75 M$_{\odot}$, encompassing the Si/O interface for many of our models, though the trends between single-star and binary-stripped star compactness remain largely unaffected for $M$ = 2-3 M$_{\odot}$ (see also \citealt{laplace21}). However, for the models studied, shock revival is determined prior to accretion of the material exterior to two solar masses, so compactness of the interior profile is more salient to our discussion. 
Generally, binary-stripped models tend to have a lower compactness than their corresponding single-star progenitors and a smaller compactness correlates with an earlier explosion in our models (we note the exceptions below). Additionally, the early accretion of a sharp Si/O density interface promotes explodability \citep{vartanyan2018a} and corresponds to a smaller compactness (due to the prompt and sharp density drop). 

For all progenitor pairs except the 17-, 18-M$_{\odot}$ models, less compact progenitors are more explodable, even where the Si/O interface may be less pronounced. Note that earlier studies have found no correlation between compactness and explodability across progenitor mass (e.g, \citealt{vartanyan2018a,radice2017b,burrows_2019,burrows_2019b,summa2016}, but see \citealt{ott2018_rel}), or a scattered correlation with higher compactness but disfavoring very high compactness \citep{2014ApJ...783...10S,2018ApJ...860...93S,2015PASJ...67..107N,ott2018_rel}. 
Compactness does correlate with certain properties of the explosion, with a higher compactness yielding higher neutrino energies and luminosities and a higher accretion rate, as well as a higher binding energy of the stellar envelope, and perhaps can serve to distinguish remnant neutron stars from black holes. However, the interplay between accretion, luminosity and explosion outcome is nuanced (see our discussion of the critical condition below) and hence a monotonic relation between compactness and explosion outcome is not expected \citep{oconnor2013}. Our conclusion is not at odds with this $-$ we emphasize that the correlation of the relative compactness with explodability here only holds for models with the same ZAMS mass, but different (single vs binary-stripped) post-main sequence evolution. We find no ``absolute compactness'' that delineates explosion from non-explosion, and a less compact model may fail to explode while a different ZAMS mass model with a higher compactness may. The results are summarized in Table\,\ref{sn_tab}, where we show the compactness, explosion outcome, the final proto-neutron star mass, and the final helium core mass for all 11 pairs of models presented here.

\subsection{Exceptions}
Contrary to the general trend of donor stars in binaries being more easily exploded than single stars of the same initial mass, we find that the 12.1- and 21-M$_{\odot}$, as well as the 17- and 18-M$_{\odot}$ single star progenitors are more explodable. The former pair can be explained by merger of the Si/O shells, resulting in a steeper Si/O interface and a less compact progenitor more conducive to explosion. The 12.1- and 21-M$_{\odot}$ single-star progenitors are indeed less compact, and more explodable, than the corresponding binary-stripped models. We find that the shell mergers happen stochastically and are not a hallmark difference of the single vs binary evolution (see \citealt{laplace21}).

For the 17- and 18-M$_{\odot}$ progenitors, we find a contrasting behavior. The single-star progenitors have a higher compactness parameter than their stripped counterparts, with Si/O interfaces further out. Surprisingly, these more compact models are more explodable than their stripped counterparts. We explore why the explosion trend is reversed for these models with a discussion of the critical condition for explosion below.

\begin{figure*}
    \centering
    \includegraphics[width=0.48\textwidth]{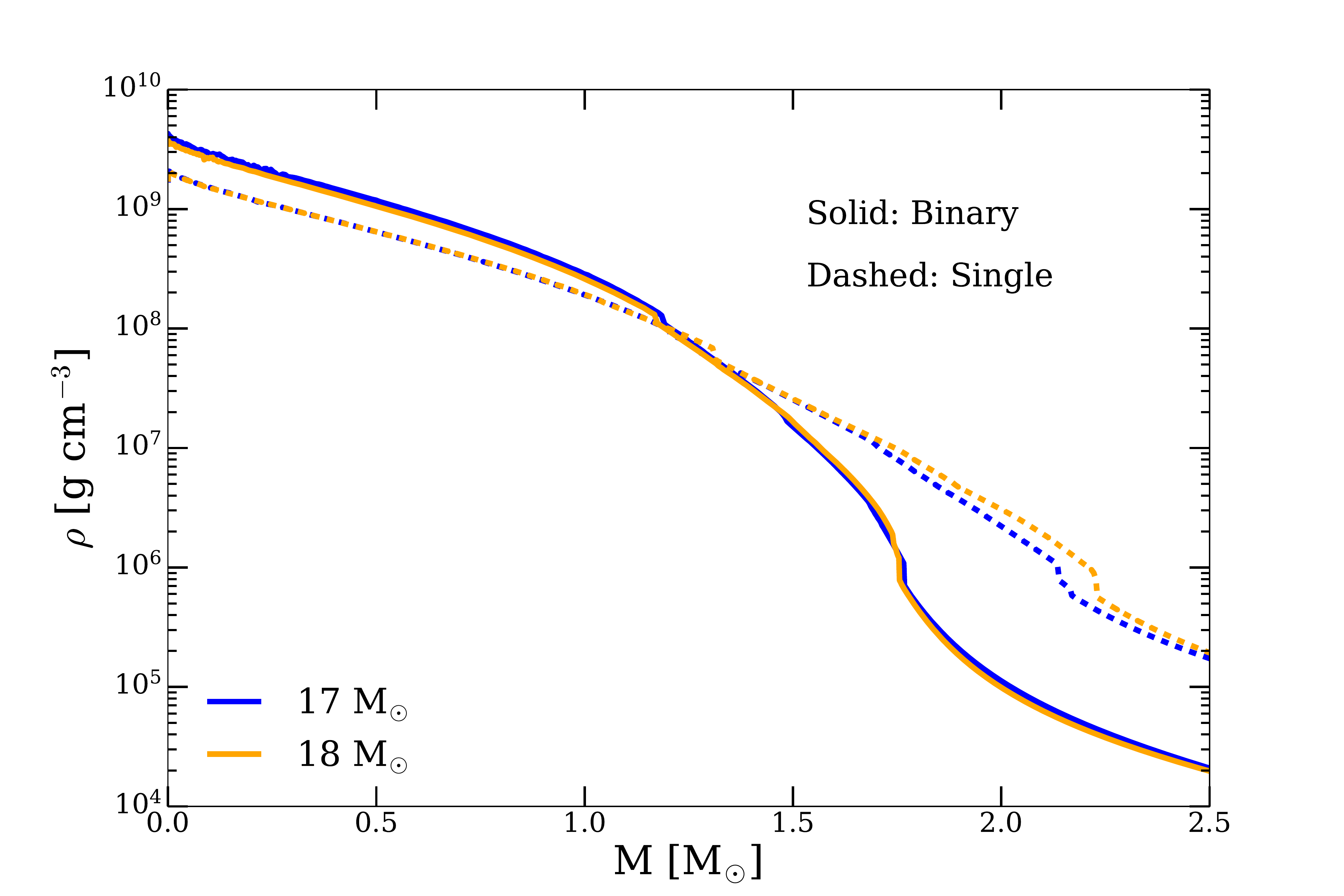}
    \includegraphics[width=0.48\textwidth]{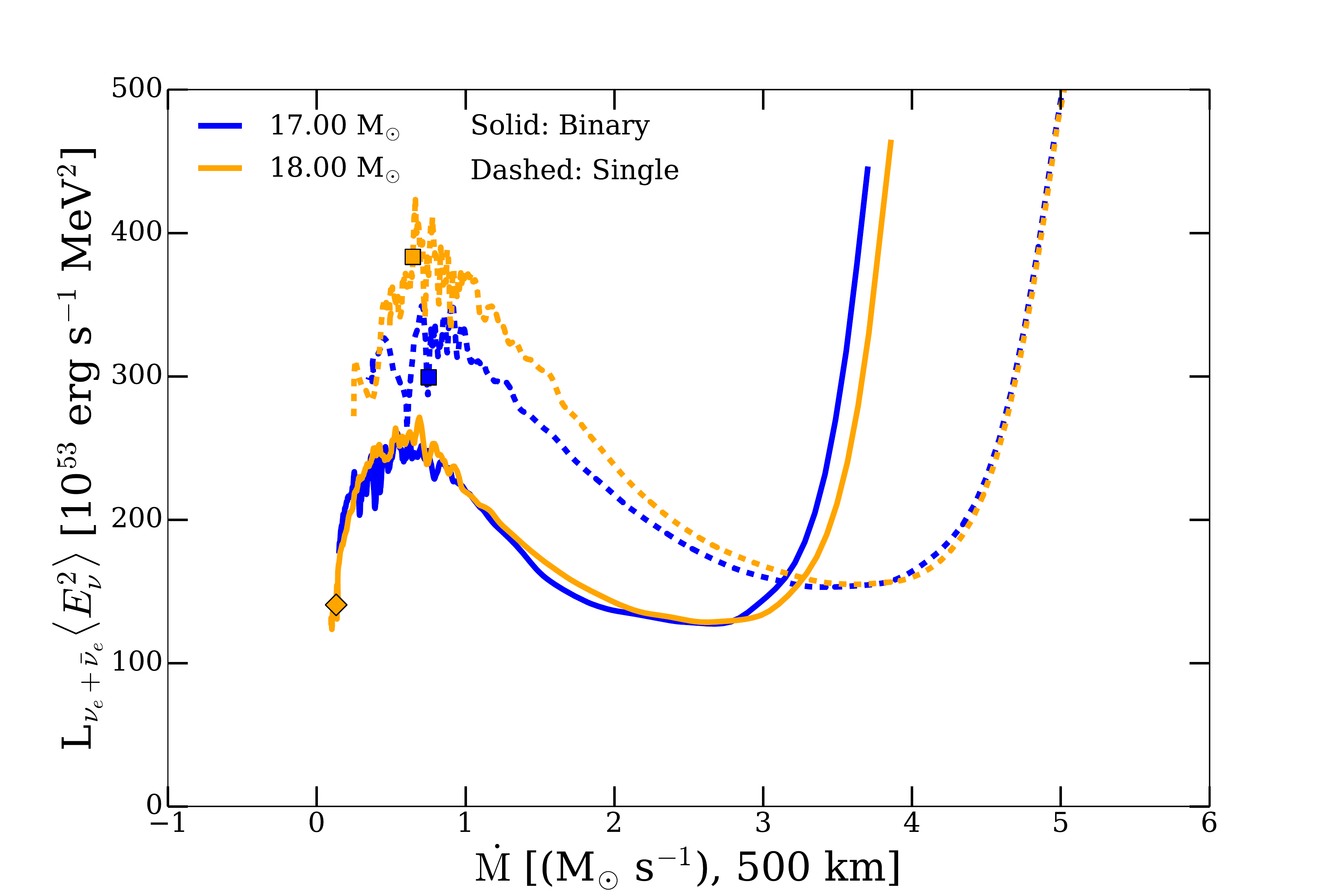}
	\caption{\textbf{Left}: We illustrate the density profiles of the 17- and 18-M$_{\odot}$ progenitors for the both the binary-stripped (solid) and single stars (dashed). Note that the more compact single-stars are more explodable, and also how similar the binary-stripped density profiles are for these two models. \textbf{Right}: We show the luminosity-accretion tracks for the 17-(orange) and 18-M$_{\odot}$ progenitors (red). We illustrate the neutrino luminosity weighted by the RMS neutrino energy (at 500 km and in 10$^{53}$ erg s$^{-1}$, summing over electron neutrinos and anti-neutrinos) as a function of accretion rate (at 500 km, in M$_{\odot}$ s$^{-1}$). The solid lines indicate the binary-stripped progenitors and the dashed the single-star progenitors. The stars evolve leftward along the illustrated tracks. The colored squares show when the single-star models explode, and the colored diamond when the stripped 18-M$_{\odot}$ model explodes. The stripped 17-M$_{\odot}$ progenitor does not explode. We see that the single-stars have a higher RMS-energy weighted neutrino luminosity for a given accretion rate than their stripped counterparts. The stripped 18-M$_{\odot}$ progenitor has a slightly higher neutrino luminosity than the stripped 17-M$_{\odot}$ prior to explosion, and the former explodes but not the latter, despite having nearly identical profiles. Near the onset of explosion, the dynamics become turbulent, as evidenced by the variations in the plotted quantities.}
	\label{fig:crit}
\end{figure*}

\subsection{Luminosity-Accretion Tracks and Criticality}
We explore the proximity to a critical condition for explosion. The critical curve \citep{burrowsgoshy1993,summa2016} quantifies the competing but coupled effects of accretion and neutrino luminosity in driving explosion. Accretion provides a ram pressure exterior to the shock which the shock must overcome to produce an explosion, while simultaneously providing an accretion-powered luminosity that contributes to neutrino heating of the stalled shock. In Fig.\,\ref{fig:crit}, we illustrate luminosity$-$accretion rate tracks for the enigmatic 17- and 18-M$_{\odot}$ models. The two single-star progenitors have a higher luminosity for a given accretion rate than the binary-stripped counterparts. The single-star progenitors for these two models have a shallower density profile that provides a higher accretion rate and a higher accretion luminosity conducive to explosion. Their subsequent explosion can be explained by their proximity to this critical condition for explosion. In addition, we find that, prior to explosion, the binary-stripped 18-M$_{\odot}$ progenitor has a slightly higher neutrino luminosity than the 
17-M$_{\odot}$ binary-stripped model, and only the former explodes despite having quite similar progenitor profiles with the latter. This highlights how sensitive explosion outcome can be to small differences in this mass range. 

However, the critical condition does not sufficiently explain explodability for other models. For instance, the 21-M$_{\odot}$ stripped progenitor has a higher luminosity for a given accretion rate than its single-star counterpart, but fails to explode. Thus, compactness and criticality provide valuable and complementary insight, and neither alone necessarily provides a definitive determinant of explosion outcome. Quantifying the exact transition from non-explosion to explosion has been a subject of previous work \citep{summa2016,summa2018}, but often requires fine-tuning the studied parameters and lies outside our present focus. We draw attention only to the point that compactness and the critical condition focus on two distinct factors of explosion outcome: the density profile, including the presence of a strong Si/O interface (see Fig.\,\ref{fig:rho_M_rshock}), and the accretion/accretion-luminosity tracks, respectively. Additionally, more detailed prescriptions for predicting explosion exist (\citealt{antesonic1,antesonic2}, \citealt{int_cond}, \citealt{2016MNRAS.460..742M}, \citealt{2016ApJ...818..124E}). These parametrizations are in the context of one-dimensional spherically-symmetric explosions, dependent on the simulation results for tuning, and beyond our scope in the context of multi-dimensional simulations.

Progenitors that are sufficiently less compact, or in our models, analogously field a prominent Si/O or equivalent compositional interface, explode regardless of their proximity to criticality. Those that are not need to satisfy the critical condition. Our results illustrate the following trend for explosion outcome by progenitor: amongst lower mass progenitors (10$-$15-M$_{\odot}$), with steeper density profiles, the less compact model is more explodable. Higher mass progenitors (17-, 18-M$_{\odot}$ single-star progenitors, and both single-star and stripped profiles for the 20-M${\odot}$ progenitors) have shallower density profiles and Si/O interfaces located further out.  Here, sustained accretion powers a sufficient neutrino luminosity for the star to explode, with the less compact star again being more explodable. For intermediate mass stars (e.g., the 17-, 18-M$_{\odot}$ binaries), the Si/O interface lies between $\sim$1.7$-$1.9 M$_{\odot}$ and is too far to be accreted for prompt shock revival, while the density profile is not shallow enough to maintain high persistent accretion to promote later shock revival. These models do not satisfy either criterion for explodability: a prominent interface or the critical condition.  

It is difficult to find a-priori indicators of explosion outcome from progenitor dependence alone, but we attempt to provide a broad categorization across progenitor mass here. No single parametrization is sufficient to capture or predict the complex nature of explosion, but jointly studying the compactness and the proximity to the critical condition span the range of possible outcomes. We emphasize that the phase space of interest for explosion outcome is really the density profile, $\rho(M)-M$, and not simply the ZAMS or He core mass of the star. A simple parametrization or explosion criterion eludes our, and the broader supernova community's, efforts.

\subsection{Explosion Diagnostics}
Explodability, and explosion timing, provides a diagnostic of the explosion properties, including remnant masses, nucleosynthetic yields, and energetics. For instance, early work (\citealt{2021Natur.589...29B}) suggests that perhaps stars that explode later may do so more energetically. In this study, this is further buttressed by the fact that the less-explodable single-stars also have a heavier mantle to sustain higher accretion energies. 

Here, we find a distribution of baryonic proto-neutron star (PNS) masses spanning 1.3 to 2.2 M$_{\odot}$ (see Table\,\ref{sn_tab}) at the end of our simulations, higher than the values found in \cite{ertl2020} for similar helium core masses. Just after core bounce, the PNS masses are close to the Chandrasekhar mass of 1.3 M$_{\odot}$, and the divergence of behavior afterwards is a function of the accretion density profile. The heavier progenitors accrete almost as much as one solar mass within the first second of core bounce. In many of our models, the PNS masses are still growing. Categorically, the binary-stripped stars, being less compact, yield PNS masses a few tenths of a solar mass smaller than their single-star pairs of the same initial mass. The exceptions, as expected, are the 12.1 and 21-M$_{\odot}$ progenitors, for which the single star progenitors are less compact and yield a smaller PNS mass. Surprisingly, both the 17- and 18-M${_\odot}$ binary-stripped progenitors produce less massive PNS, despite being less explodable than their single-star counterparts, suggesting that progenitor density profile and compactness may be additionally critical to determining PNS mass than simply the explosion time alone.

The 18- and 20-M$_{\odot}$ stripped-star progenitors both yield PNS masses over 2 M$_{\odot}$, and both successfully explode as well, providing simulation insight into the upper mass limits of neutron stars. After explosion within the first second, multiple solar masses of material remain in the stellar envelope to be ultimately accreted and, in smaller part, blown out as a wind. The possible existence of a mass gap \citep{bailyn,Ozel+2010,farr} between the most massive neutron stars (e.g., \citealt{cromartie}) and the least massive black holes (e.g., \citealt{2020arXiv200612611T}), lying between $\sim$2.5 to 5 M$_{\odot}$ could be populated by fallback accretion onto the PNS remnants of stripped stars in binaries \citep{Fryer+2012,Woosley_2020,2021A&A...645A...5S}. Late time simulations of CCSNe that encompass mass fallback are lacking, as is our understanding of the details of fallback accretion. However, we can approximate fallback by assuming the morphology of the shock (e.g., bottom right panel of Fig.\,\ref{fig:typical}) is sustained and we can estimate a volume-filling fraction of the expanding shock. At 5000 km, this is $\sim$20\%, suggesting that more than half of the envelope will accrete at later times. Indeed, we only require sustained accretion of less than half the helium core mass of binary-stripped stars (see Table\,\ref{sn_tab}), or around 1-2 solar masses, to populate this mass gap. We emphasize again that this is just a cursory estimate, and detailed, long-term 3D simulations are necessary.

The successfully exploding models present a typical diagnostic explosion energy of $\sim$0.2 Bethe (1 Bethe $\equiv$ 10$^{51}$ erg) within the first second. The more massive single 17-,18-M$_{\odot}$ progenitors have energies of upwards 0.4 Bethe, and both the single and binary-stripped 20-$M_{\odot}$ progenitors have diagnostic energies of more than 0.5 Bethe. However, correcting for the gravitational binding energy of the material exterior to our grid, we find that all of our models are still gravitationally-bound, albeit tenuously. This is to be expected for a short simulation, spanning less than one second post-bounce. Furthermore, many of the models show significant early energy growth rates (consistent with \citealt{burrows_2019b,2021Natur.589...29B}), suggesting that longer simulations are necessary to see the final explosion energies for many of these models, as well as the remnant PNS masses, nucleosynthetic yields, and possible remnant kick velocities.

Additionally, although some single-star models (such as the 18-, 20-M$_{\odot}$ models) have higher diagnostic explosion energies, all binary-stripped models have more weakly bound and less massive envelopes. Thus, correcting for the gravitational overburden of the envelope, all of our exploding binary-stripped models have higher net explosion energies than the corresponding single stars, at least within the first second of simulation. However, our preliminary results seem to suggest that single stars have faster energy growth rates than their binary-stripped pairs, which we attribute to a larger accreting mass. This is at odds with \cite{2021A&A...645A...5S}, who find a higher accreting mass and higher energies in their binary-stripped models using a parametric supernova code. Resolution of this necessitates longer simulations in self-consistent 3D. In addition, we find that single stars tend to have higher gravitational-ave energies, potentially due to more massive envelopes ultimately yielding greater turbulent energy impinging on the PNS (\citealt{radice2019,2019MNRAS.489.2227V,2020ApJ...901..108V}).

Though the helium core mass correlates with the core compactness, compactness itself does not provide a global metric for explosion outcome. Rather, we find that core compactness serves as a relative metric for explodability for progenitors of the same ZAMS mass but different evolutionary channels. Insofar as the helium core mass correlates with the compactness, which we find is a useful relative metric for explodability of binary-stripped and single star pairs, we find a weak and scattered correlation with more massive helium cores exploding later and with higher PNS masses. However, the helium core extends too far in mass to significantly affect the early nature of shock revival, whereas the interior density profile is critical to a successful explosion.

\section{Conclusions}
\label{sec:conc}

In this study, we presented a comparison of binary-stripped stars with single-star counterparts of the same initial mass. Compared to current literature, our work provides an improvement along two avenues $-$ we follow a self-consistent evolution of both the pre-collapse progenitors, including binary interactions as our initial condition and the subsequent explosion outcome. We illustrated that binary-stripped progenitors typically are less compact and tend to be more explodable (see also \citealt{woosley2019,2021A&A...645A...5S}) than single stars with the same initial mass.  The latter remains to be confirmed $-$ our detailed simulations post-bounce and stellar evolution pre-bounce are hostage to uncertainties in the understanding of stellar theory, including the development of convection \citep{1987A&A...188...49R,2020MNRAS.493.4333R}, nuclear burning reactions \citep{farmer_variations_2016}, winds \citep{2017A&A...603A.118R},  overshooting \citep{2019MNRAS.484.3921D}, and the detailed of role of neutrino microphysics. However, these initial results are promising to resolve possible discrepancy between the stellar mass function and the rate of supernovae, as well as populate the possible neutron-star black hole mass gap. The higher explodability of binary-stripped stars may also help explain the large fraction of stripped-envelope supernovae. We expect the nucleosynthesis of single and binary-stripped stars to be different (Farmer et al. in prep.) due to the systematic differences in composition at the onset of core collapse (\citealt{laplace21}). If binary-stripped stars are more explodable, then these differences in nucleosynthesis are even more important.

Stars of similar initial masses can have very different density profiles (e.g., \citealt{2014ApJ...783...10S,2018ApJ...860...93S}) and hence different explosion outcomes. Here, we have seen stars of different masses, e.g. the 17-, 18-M$_{\odot}$ models, with very similar density profiles but qualitatively different explosion outcomes (explosion vs. no explosion). We find only a weak, scattered dependence on explosion outcome with the helium core mass. 

Earlier work on binary progenitors of CCSNe have either evolved only until the formation of a carbon-oxygen core, or evolved an isolated carbon-oxygen core until collapse \citep{patton2020}. Additionally, these studies have either artificially inducing explosion through parametrized heating (\citealt{woosley2019,2021A&A...645A...5S}), or have used prescriptive formulae to predict explosion outcome, introducing uncertainty in both the progenitor profile and its final fate. Here we present a self-consistent study of explosion outcome of stripped-models with their single-star counterparts. We find that the progenitor density profile is critical to explosion outcome. The sensitivity of the explosion to details necessitates a through study of the physical uncertainties in the progenitor models $-$ the initial conditions for core-collapse simulations $-$ before we can confidently claim resolution of the core-collapse problem. Recent results would indicate we have reached a tipping point where uncertainties in the stellar evolution models dwarf uncertainties in the neutrino-heated explosion of core-collapse supernovae.

\section*{Acknowledgments}
We acknowledge Tony Piro, Stephen Justham, Robert Farmer, and Daniel Kasen for valuable discussion. DV and AB acknowledge support from the U.S. Department of Energy Office of Science and the Office
of Advanced Scientific Computing Research via the Scientific Discovery
through Advanced Computing (SciDAC4) program and Grant DE-SC0018297
(subaward 00009650) and support from the U.S. NSF under Grants AST-1714267
and PHY-1804048 (the latter via the Max-Planck/Princeton Center (MPPC) for Plasma Physics). EL and SdM acknowledge the European Union’s Horizon 2020 research and innovation program from the European Research Council (ERC, grant agreement No. 715063) and the Netherlands Organisation for Scientific Research (NWO) as part of the Vidi research program BinWaves with project number 639.042.728. YG acknowledges the funding from the Alvin E. Nashman fellowship for Theoretical Astrophysics. The authors employed computational resources provided by the TIGRESS high
performance computer center at Princeton University, which is jointly
supported by the Princeton Institute for Computational Science and
Engineering (PICSciE) and the Princeton University Office of Information
Technology.
    

\bibliographystyle{aasjournal} 
\bibliography{References}

\begin{thebibliography}{}
\expandafter\ifx\csname natexlab\endcsname\relax\def\natexlab#1{#1}\fi
\providecommand{\url}[1]{\href{#1}{#1}}
\providecommand{\dodoi}[1]{doi:~\href{http://doi.org/#1}{\nolinkurl{#1}}}
\providecommand{\doeprint}[1]{\href{http://ascl.net/#1}{\nolinkurl{http://ascl.net/#1}}}
\providecommand{\doarXiv}[1]{\href{https://arxiv.org/abs/#1}{\nolinkurl{https://arxiv.org/abs/#1}}}

\bibitem[{{Almeida} {et~al.}(2017){Almeida}, {Sana}, {Taylor}, {Barb{\'a}},
  {Bonanos}, {Crowther}, {Damineli}, {de Koter}, {de Mink}, {Evans}, {Gieles},
  {Grin}, {H{\'e}nault-Brunet}, {Langer}, {Lennon}, {Lockwood}, {Ma{\'{\i}}z
  Apell{\'a}niz}, {Moffat}, {Neijssel}, {Norman}, {Ram{\'{\i}}rez-Agudelo},
  {Richardson}, {Schootemeijer}, {Shenar}, {Soszy{\'n}ski}, {Tramper}, \&
  {Vink}}]{almeida:17}
{Almeida}, L.~A., {Sana}, H., {Taylor}, W., {et~al.} 2017, \aap, 598, A84,
  \dodoi{10.1051/0004-6361/201629844}

\bibitem[{Asplund {et~al.}(2009)Asplund, Grevesse, Sauval, \&
  Scott}]{asplund_chemical_2009}
Asplund, M., Grevesse, N., Sauval, A.~J., \& Scott, P. 2009, Annual Review of
  Astronomy and Astrophysics, 47, 481,
  \dodoi{10.1146/annurev.astro.46.060407.145222}

\bibitem[{{Bailyn} {et~al.}(1998){Bailyn}, {Jain}, {Coppi}, \&
  {Orosz}}]{bailyn}
{Bailyn}, C.~D., {Jain}, R.~K., {Coppi}, P., \& {Orosz}, J.~A. 1998, \apj, 499,
  367, \dodoi{10.1086/305614}

\bibitem[{{Burrows} \& {Goshy}(1993)}]{burrowsgoshy1993}
{Burrows}, A., \& {Goshy}, J. 1993, \apjl, 416, L75, \dodoi{10.1086/187074}

\bibitem[{{Burrows} {et~al.}(2019){Burrows}, {Radice}, \&
  {Vartanyan}}]{burrows_2019}
{Burrows}, A., {Radice}, D., \& {Vartanyan}, D. 2019, \mnras, 485, 3153,
  \dodoi{10.1093/mnras/stz543}

\bibitem[{Burrows {et~al.}(2019)Burrows, Radice, Vartanyan, Nagakura, Skinner,
  \& Dolence}]{burrows_2019b}
Burrows, A., Radice, D., Vartanyan, D., {et~al.} 2019, Monthly Notices of the
  Royal Astronomical Society, 491, 2715–2735, \dodoi{10.1093/mnras/stz3223}

\bibitem[{{Burrows} \& {Vartanyan}(2021)}]{2021Natur.589...29B}
{Burrows}, A., \& {Vartanyan}, D. 2021, \nat, 589, 29,
  \dodoi{10.1038/s41586-020-03059-w}

\bibitem[{Burrows {et~al.}(2018)Burrows, Vartanyan, Dolence, Skinner, \&
  Radice}]{burrows2018}
Burrows, A., Vartanyan, D., Dolence, J.~C., Skinner, M.~A., \& Radice, D. 2018,
  Space Science Reviews, 214, 33, \dodoi{10.1007/s11214-017-0450-9}

\bibitem[{{Claeys} {et~al.}(2011){Claeys}, {de Mink}, {Pols}, {Eldridge}, \&
  {Baes}}]{Claeys+2011}
{Claeys}, J.~S.~W., {de Mink}, S.~E., {Pols}, O.~R., {Eldridge}, J.~J., \&
  {Baes}, M. 2011, \aap, 528, A131

\bibitem[{{Couch} \& {Ott}(2013)}]{2013ApJ...778L...7C}
{Couch}, S.~M., \& {Ott}, C.~D. 2013, \apjl, 778, L7,
  \dodoi{10.1088/2041-8205/778/1/L7}

\bibitem[{{Cromartie} {et~al.}(2020){Cromartie}, {Fonseca}, {Ransom},
  {Demorest}, {Arzoumanian}, {Blumer}, {Brook}, {DeCesar}, {Dolch}, {Ellis},
  {Ferdman}, {Ferrara}, {Garver-Daniels}, {Gentile}, {Jones}, {Lam}, {Lorimer},
  {Lynch}, {McLaughlin}, {Ng}, {Nice}, {Pennucci}, {Spiewak}, {Stairs},
  {Stovall}, {Swiggum}, \& {Zhu}}]{cromartie}
{Cromartie}, H.~T., {Fonseca}, E., {Ransom}, S.~M., {et~al.} 2020, Nature
  Astronomy, 4, 72, \dodoi{10.1038/s41550-019-0880-2}

\bibitem[{{Davis} {et~al.}(2019){Davis}, {Jones}, \&
  {Herwig}}]{2019MNRAS.484.3921D}
{Davis}, A., {Jones}, S., \& {Herwig}, F. 2019, \mnras, 484, 3921,
  \dodoi{10.1093/mnras/sty3415}

\bibitem[{{De} {et~al.}(2018){De}, {Kasliwal}, {Ofek}, {Moriya}, {Burke},
  {Cao}, {Cenko}, {Doran}, {Duggan}, {Fender}, {Fransson}, {Gal-Yam}, {Horesh},
  {Kulkarni}, {Laher}, {Lunnan}, {Manulis}, {Masci}, {Mazzali}, {Nugent},
  {Perley}, {Petrushevska}, {Piro}, {Rumsey}, {Sollerman}, {Sullivan}, \&
  {Taddia}}]{de2018}
{De}, K., {Kasliwal}, M.~M., {Ofek}, E.~O., {et~al.} 2018, Science, 362, 201,
  \dodoi{10.1126/science.aas8693}

\bibitem[{{de~Mink} {et~al.}(2008){de~Mink}, {Pols}, \&
  {Yoon}}]{2008AIPC..990..230D}
{de~Mink}, S.~E., {Pols}, O.~R., \& {Yoon}, S.~C. 2008, in American Institute
  of Physics Conference Series, Vol. 990, First Stars III, ed. B.~W. {O'Shea}
  \& A.~{Heger}, 230--232, \dodoi{10.1063/1.2905549}

\bibitem[{Eldridge {et~al.}(2013)Eldridge, Fraser, Smartt, Maund, \&
  Crockett}]{eldridge_death_2013}
Eldridge, J.~J., Fraser, M., Smartt, S.~J., Maund, J.~R., \& Crockett, R.~M.
  2013, Monthly Notices of the Royal Astronomical Society, 436, 774,
  \dodoi{10.1093/mnras/stt1612}

\bibitem[{{Eldridge} {et~al.}(2008){Eldridge}, {Izzard}, \&
  {Tout}}]{Eldridge+2008}
{Eldridge}, J.~J., {Izzard}, R.~G., \& {Tout}, C.~A. 2008, \mnras, 384, 1109

\bibitem[{Eldridge {et~al.}(2017)Eldridge, Stanway, Xiao, McClelland, Taylor,
  Ng, Greis, \& Bray}]{eldridge_binary_2017}
Eldridge, J.~J., Stanway, E.~R., Xiao, L., {et~al.} 2017, Publications of the
  Astronomical Society of Australia, 34, e058, \dodoi{10.1017/pasa.2017.51}

\bibitem[{{Eldridge} {et~al.}(2018){Eldridge}, {Xiao}, {Stanway}, {Rodrigues},
  \& {Guo}}]{eldridge:18}
{Eldridge}, J.~J., {Xiao}, L., {Stanway}, E.~R., {Rodrigues}, N., \& {Guo},
  N.~Y. 2018, \pasa, 35, 49, \dodoi{10.1017/pasa.2018.47}

\bibitem[{{Ertl} {et~al.}(2016){Ertl}, {Janka}, {Woosley}, {Sukhbold}, \&
  {Ugliano}}]{2016ApJ...818..124E}
{Ertl}, T., {Janka}, H.-T., {Woosley}, S.~E., {Sukhbold}, T., \& {Ugliano}, M.
  2016, \apj, 818, 124, \dodoi{10.3847/0004-637X/818/2/124}

\bibitem[{{Ertl} {et~al.}(2020){Ertl}, {Woosley}, {Sukhbold}, \&
  {Janka}}]{ertl2020}
{Ertl}, T., {Woosley}, S.~E., {Sukhbold}, T., \& {Janka}, H.~T. 2020, \apj,
  890, 51, \dodoi{10.3847/1538-4357/ab6458}

\bibitem[{Farmer {et~al.}(2016)Farmer, Fields, Petermann, Dessart, Cantiello,
  Paxton, \& Timmes}]{farmer_variations_2016}
Farmer, R., Fields, C.~E., Petermann, I., {et~al.} 2016, The Astrophysical
  Journal Supplement Series, 227, 22, \dodoi{10.3847/1538-4365/227/2/22}

\bibitem[{{Farr} {et~al.}(2011){Farr}, {Sravan}, {Cantrell}, {Kreidberg},
  {Bailyn}, {Mandel}, \& {Kalogera}}]{farr}
{Farr}, W.~M., {Sravan}, N., {Cantrell}, A., {et~al.} 2011, \apj, 741, 103,
  \dodoi{10.1088/0004-637X/741/2/103}

\bibitem[{Fox {et~al.}(2014)Fox, Azalee~Bostroem, Van~Dyk, Filippenko,
  Fransson, Matheson, Cenko, Chandra, Dwarkadas, Li, Parker, \&
  Smith}]{fox_uncovering_2014}
Fox, O.~D., Azalee~Bostroem, K., Van~Dyk, S.~D., {et~al.} 2014, The
  Astrophysical Journal, 790, 17, \dodoi{10.1088/0004-637X/790/1/17}

\bibitem[{{Fryer}(1999)}]{fryer1999}
{Fryer}, C.~L. 1999, \apj, 522, 413, \dodoi{10.1086/307647}

\bibitem[{{Fryer} {et~al.}(2012){Fryer}, {Belczynski}, {Wiktorowicz},
  {Dominik}, {Kalogera}, \& {Holz}}]{Fryer+2012}
{Fryer}, C.~L., {Belczynski}, K., {Wiktorowicz}, G., {et~al.} 2012, \apj, 749,
  91

\bibitem[{{Gilkis} {et~al.}(2019){Gilkis}, {Vink}, {Eldridge}, \&
  {Tout}}]{Gilkis+2019}
{Gilkis}, A., {Vink}, J.~S., {Eldridge}, J.~J., \& {Tout}, C.~A. 2019, \mnras,
  486, 4451

\bibitem[{{Glas} {et~al.}(2019){Glas}, {Just}, {Janka}, \&
  {Obergaulinger}}]{glas2019}
{Glas}, R., {Just}, O., {Janka}, H.~T., \& {Obergaulinger}, M. 2019, \apj, 873,
  45, \dodoi{10.3847/1538-4357/ab0423}

\bibitem[{{G{\"o}tberg} {et~al.}(2017){G{\"o}tberg}, {de Mink}, \&
  {Groh}}]{gotberg:17}
{G{\"o}tberg}, Y., {de Mink}, S.~E., \& {Groh}, J.~H. 2017, \aap, 608, A11,
  \dodoi{10.1051/0004-6361/201730472}

\bibitem[{Kippenhahn \& Weigert(1967)}]{kippenhahn_entwicklung_1967}
Kippenhahn, R., \& Weigert, A. 1967, Zeitschrift fur Astrophysik, 65, 251.
\newblock \url{https://ui.adsabs.harvard.edu/#abs/1967ZA.....65..251K/abstract}

\bibitem[{{Klencki} {et~al.}(2020){Klencki}, {Nelemans}, {Istrate}, \&
  {Pols}}]{Klencki+2020}
{Klencki}, J., {Nelemans}, G., {Istrate}, A.~G., \& {Pols}, O. 2020, \aap, 638,
  A55

\bibitem[{{Kuroda} {et~al.}(2020){Kuroda}, {Arcones}, {Takiwaki}, \&
  {Kotake}}]{kuroda2020}
{Kuroda}, T., {Arcones}, A., {Takiwaki}, T., \& {Kotake}, K. 2020, arXiv
  e-prints, arXiv:2003.02004.
\newblock \doarXiv{2003.02004}

\bibitem[{{Langer}(1989)}]{1989A&A...210...93L}
{Langer}, N. 1989, \aap, 210, 93

\bibitem[{{Laplace} {et~al.}(2021){Laplace}, {Justham}, {Renzo}, {G{\"o}tberg},
  {Farmer}, {Vartanyan}, \& {de Mink}}]{laplace21}
{Laplace}, E., {Justham}, S., {Renzo}, M., {et~al.} 2021, arXiv e-prints,
  arXiv:2102.05036.
\newblock \doarXiv{2102.05036}

\bibitem[{Li {et~al.}(2011)Li, Leaman, Chornock, Filippenko, Poznanski,
  Ganeshalingam, Wang, Modjaz, Jha, Foley, \& Smith}]{li_nearby_2011}
Li, W., Leaman, J., Chornock, R., {et~al.} 2011, Monthly Notices of the Royal
  Astronomical Society, 412, 1441, \dodoi{10.1111/j.1365-2966.2011.18160.x}

\bibitem[{Lyman {et~al.}(2016)Lyman, Bersier, James, Mazzali, Eldridge, Fraser,
  \& Pian}]{lyman_bolometric_2016}
Lyman, J.~D., Bersier, D., James, P.~A., {et~al.} 2016, Monthly Notices of the
  Royal Astronomical Society, 457, 328, \dodoi{10.1093/mnras/stv2983}

\bibitem[{{Marek} {et~al.}(2006){Marek}, {Dimmelmeier}, {Janka}, {M{\"u}ller},
  \& {Buras}}]{marek2006}
{Marek}, A., {Dimmelmeier}, H., {Janka}, H.-T., {M{\"u}ller}, E., \& {Buras},
  R. 2006, \aap, 445, 273, \dodoi{10.1051/0004-6361:20052840}

\bibitem[{{Mason} {et~al.}(2009){Mason}, {Hartkopf}, {Gies}, {Henry}, \&
  {Helsel}}]{mason:09}
{Mason}, B.~D., {Hartkopf}, W.~I., {Gies}, D.~R., {Henry}, T.~J., \& {Helsel},
  J.~W. 2009, \aj, 137, 3358, \dodoi{10.1088/0004-6256/137/2/3358}

\bibitem[{Maund {et~al.}(2004)Maund, Smartt, Kudritzki, Podsiadlowski, \&
  Gilmore}]{maund_massive_2004}
Maund, J.~R., Smartt, S.~J., Kudritzki, R.~P., Podsiadlowski, P., \& Gilmore,
  G.~F. 2004, Nature, 427, 129, \dodoi{10.1038/nature02161}

\bibitem[{{M{\"u}ller} {et~al.}(2018){M{\"u}ller}, {Gay}, {Heger}, {Tauris}, \&
  {Sim}}]{mueller:18}
{M{\"u}ller}, B., {Gay}, D.~W., {Heger}, A., {Tauris}, T.~M., \& {Sim}, S.~A.
  2018, \mnras, 479, 3675, \dodoi{10.1093/mnras/sty1683}

\bibitem[{{M{\"u}ller} {et~al.}(2016){M{\"u}ller}, {Heger}, {Liptai}, \&
  {Cameron}}]{2016MNRAS.460..742M}
{M{\"u}ller}, B., {Heger}, A., {Liptai}, D., \& {Cameron}, J.~B. 2016, \mnras,
  460, 742, \dodoi{10.1093/mnras/stw1083}

\bibitem[{{M{\"u}ller} {et~al.}(2019){M{\"u}ller}, {Tauris}, {Heger},
  {Banerjee}, {Qian}, {Powell}, {Chan}, {Gay}, \& {Langer}}]{mueller:19}
{M{\"u}ller}, B., {Tauris}, T.~M., {Heger}, A., {et~al.} 2019, \mnras, 484,
  3307, \dodoi{10.1093/mnras/stz216}

\bibitem[{{Murphy} \& {Dolence}(2017)}]{int_cond}
{Murphy}, J.~W., \& {Dolence}, J.~C. 2017, \apj, 834, 183,
  \dodoi{10.3847/1538-4357/834/2/183}

\bibitem[{{Nagakura} {et~al.}(2019){Nagakura}, {Sumiyoshi}, \&
  {Yamada}}]{nagakura2019}
{Nagakura}, H., {Sumiyoshi}, K., \& {Yamada}, S. 2019, arXiv e-prints,
  arXiv:1906.10143.
\newblock \doarXiv{1906.10143}

\bibitem[{{Nakamura} {et~al.}(2015){Nakamura}, {Takiwaki}, {Kuroda}, \&
  {Kotake}}]{2015PASJ...67..107N}
{Nakamura}, K., {Takiwaki}, T., {Kuroda}, T., \& {Kotake}, K. 2015, \pasj, 67,
  107, \dodoi{10.1093/pasj/psv073}

\bibitem[{{O'Connor} \& {Ott}(2011)}]{2011ApJ...730...70O}
{O'Connor}, E., \& {Ott}, C.~D. 2011, \apj, 730, 70,
  \dodoi{10.1088/0004-637X/730/2/70}

\bibitem[{{O'Connor} \& {Ott}(2013)}]{oconnor2013}
---. 2013, \apj, 762, 126, \dodoi{10.1088/0004-637X/762/2/126}

\bibitem[{{O'Connor} \& {Couch}(2018)}]{oconnor_couch2018b}
{O'Connor}, E.~P., \& {Couch}, S.~M. 2018, \apj, 865, 81,
  \dodoi{10.3847/1538-4357/aadcf7}

\bibitem[{{Ott} {et~al.}(2018){Ott}, {Roberts}, {da Silva Schneider}, {Fedrow},
  {Haas}, \& {Schnetter}}]{ott2018_rel}
{Ott}, C.~D., {Roberts}, L.~F., {da Silva Schneider}, A., {et~al.} 2018, \apjl,
  855, L3, \dodoi{10.3847/2041-8213/aaa967}

\bibitem[{{{\"O}zel} {et~al.}(2010){{\"O}zel}, {Psaltis}, {Narayan}, \&
  {McClintock}}]{Ozel+2010}
{{\"O}zel}, F., {Psaltis}, D., {Narayan}, R., \& {McClintock}, J.~E. 2010,
  \apj, 725, 1918

\bibitem[{{Patton} \& {Sukhbold}(2020)}]{patton2020}
{Patton}, R.~A., \& {Sukhbold}, T. 2020, \mnras, 499, 2803,
  \dodoi{10.1093/mnras/staa3029}

\bibitem[{Paxton {et~al.}(2011)Paxton, Bildsten, Dotter, Herwig, Lesaffre, \&
  Timmes}]{paxton_modules_2011}
Paxton, B., Bildsten, L., Dotter, A., {et~al.} 2011, The Astrophysical Journal
  Supplement Series, 192, 3, \dodoi{10.1088/0067-0049/192/1/3}

\bibitem[{Paxton {et~al.}(2013)Paxton, Cantiello, Arras, Bildsten, Brown,
  Dotter, Mankovich, Montgomery, Stello, Timmes, \&
  Townsend}]{paxton_modules_2013}
Paxton, B., Cantiello, M., Arras, P., {et~al.} 2013, The Astrophysical Journal
  Supplement Series, 208, 4, \dodoi{10.1088/0067-0049/208/1/4}

\bibitem[{Paxton {et~al.}(2015)Paxton, Marchant, Schwab, Bauer, Bildsten,
  Cantiello, Dessart, Farmer, Hu, Langer, Townsend, Townsley, \&
  Timmes}]{paxton_modules_2015}
Paxton, B., Marchant, P., Schwab, J., {et~al.} 2015, The Astrophysical Journal
  Supplement Series, 220, 15, \dodoi{10.1088/0067-0049/220/1/15}

\bibitem[{Paxton {et~al.}(2018)Paxton, Schwab, Bauer, Bildsten, Blinnikov,
  Duffell, Farmer, Goldberg, Marchant, Sorokina, Thoul, Townsend, \&
  Timmes}]{paxton_modules_2018}
Paxton, B., Schwab, J., Bauer, E.~B., {et~al.} 2018, The Astrophysical Journal
  Supplement Series, 234, 34, \dodoi{10.3847/1538-4365/aaa5a8}

\bibitem[{Paxton {et~al.}(2019)Paxton, Smolec, Schwab, Gautschy, Bildsten,
  Cantiello, Dotter, Farmer, Goldberg, Jermyn, Kanbur, Marchant, Thoul,
  Townsend, Wolf, Zhang, \& Timmes}]{paxton_modules_2019-1}
Paxton, B., Smolec, R., Schwab, J., {et~al.} 2019, The Astrophysical Journal
  Supplement Series, 243, 10, \dodoi{10.3847/1538-4365/ab2241}

\bibitem[{{Pejcha} \& {Thompson}(2012)}]{antesonic1}
{Pejcha}, O., \& {Thompson}, T.~A. 2012, \apj, 746, 106,
  \dodoi{10.1088/0004-637X/746/1/106}

\bibitem[{{Podsiadlowski} {et~al.}(1992){Podsiadlowski}, {Joss}, \&
  {Hsu}}]{Podsiadlowski+1992}
{Podsiadlowski}, P., {Joss}, P.~C., \& {Hsu}, J.~J.~L. 1992, \apj, 391, 246

\bibitem[{{Prentice} {et~al.}(2019){Prentice}, {Ashall}, {James}, {Short},
  {Mazzali}, {Bersier}, {Crowther}, {Barbarino}, {Chen}, {Copperwheat},
  {Darnley}, {Denneau}, {Elias-Rosa}, {Fraser}, {Galbany}, {Gal-Yam},
  {Harmanen}, {Howell}, {Hosseinzadeh}, {Inserra}, {Kankare}, {Karamehmetoglu},
  {Lamb}, {Limongi}, {Maguire}, {McCully}, {Olivares E}, {Piascik}, {Pignata},
  {Reichart}, {Rest}, {Reynolds}, {Rodr{\'\i}guez}, {Saario}, {Schulze},
  {Smartt}, {Smith}, {Sollerman}, {Stalder}, {Sullivan}, {Taddia}, {Valenti},
  {Vergani}, {Williams}, \& {Young}}]{2019MNRAS.485.1559P}
{Prentice}, S.~J., {Ashall}, C., {James}, P.~A., {et~al.} 2019, \mnras, 485,
  1559, \dodoi{10.1093/mnras/sty3399}

\bibitem[{{Pursiainen} {et~al.}(2018){Pursiainen}, {Childress}, {Smith},
  {Prajs}, {Sullivan}, {Davis}, {Foley}, {Asorey}, {Calcino}, {Carollo},
  {Curtin}, {D'Andrea}, {Glazebrook}, {Gutierrez}, {Hinton}, {Hoormann},
  {Inserra}, {Kessler}, {King}, {Kuehn}, {Lewis}, {Lidman}, {Macaulay},
  {M{\"o}ller}, {Nichol}, {Sako}, {Sommer}, {Swann}, {Tucker}, {Uddin},
  {Wiseman}, {Zhang}, {Abbott}, {Abdalla}, {Allam}, {Annis}, {Avila}, {Brooks},
  {Buckley-Geer}, {Burke}, {Carnero Rosell}, {Carrasco Kind}, {Carretero},
  {Castander}, {Cunha}, {Davis}, {De Vicente}, {Diehl}, {Doel}, {Eifler},
  {Flaugher}, {Fosalba}, {Frieman}, {Garc{\'\i}a-Bellido}, {Gruen}, {Gruendl},
  {Gutierrez}, {Hartley}, {Hollowood}, {Honscheid}, {James}, {Jeltema},
  {Kuropatkin}, {Li}, {Lima}, {Maia}, {Martini}, {Menanteau}, {Ogando},
  {Plazas}, {Roodman}, {Sanchez}, {Scarpine}, {Schindler}, {Smith},
  {Soares-Santos}, {Sobreira}, {Suchyta}, {Swanson}, {Tarle}, {Tucker},
  {Walker}, \& {DES Collaboration}}]{2018MNRAS.481..894P}
{Pursiainen}, M., {Childress}, M., {Smith}, M., {et~al.} 2018, \mnras, 481,
  894, \dodoi{10.1093/mnras/sty2309}

\bibitem[{{Radice} {et~al.}(2017){Radice}, {Burrows}, {Vartanyan}, {Skinner},
  \& {Dolence}}]{radice2017b}
{Radice}, D., {Burrows}, A., {Vartanyan}, D., {Skinner}, M.~A., \& {Dolence},
  J.~C. 2017, \apj, 850, 43, \dodoi{10.3847/1538-4357/aa92c5}

\bibitem[{{Radice} {et~al.}(2019){Radice}, {Morozova}, {Burrows}, {Vartanyan},
  \& {Nagakura}}]{radice2019}
{Radice}, D., {Morozova}, V., {Burrows}, A., {Vartanyan}, D., \& {Nagakura}, H.
  2019, \apj, 876, L9, \dodoi{10.3847/2041-8213/ab191a}

\bibitem[{{Raives} {et~al.}(2018){Raives}, {Couch}, {Greco}, {Pejcha}, \&
  {Thompson}}]{antesonic2}
{Raives}, M.~J., {Couch}, S.~M., {Greco}, J.~P., {Pejcha}, O., \& {Thompson},
  T.~A. 2018, ArXiv e-prints.
\newblock \doarXiv{1801.02626}

\bibitem[{{Renzini}(1987)}]{1987A&A...188...49R}
{Renzini}, A. 1987, \aap, 188, 49

\bibitem[{{Renzo} {et~al.}(2020){Renzo}, {Farmer}, {Justham}, {de Mink},
  {G{\"o}tberg}, \& {Marchant}}]{2020MNRAS.493.4333R}
{Renzo}, M., {Farmer}, R.~J., {Justham}, S., {et~al.} 2020, \mnras, 493, 4333,
  \dodoi{10.1093/mnras/staa549}

\bibitem[{{Renzo} {et~al.}(2017){Renzo}, {Ott}, {Shore}, \& {de
  Mink}}]{2017A&A...603A.118R}
{Renzo}, M., {Ott}, C.~D., {Shore}, S.~N., \& {de Mink}, S.~E. 2017, \aap, 603,
  A118, \dodoi{10.1051/0004-6361/201730698}

\bibitem[{{Roberts} {et~al.}(2016){Roberts}, {Ott}, {Haas}, {O'Connor},
  {Diener}, \& {Schnetter}}]{roberts2016}
{Roberts}, L.~F., {Ott}, C.~D., {Haas}, R., {et~al.} 2016, \apj, 831, 98,
  \dodoi{10.3847/0004-637X/831/1/98}

\bibitem[{Ryder {et~al.}(2018)Ryder, Van~Dyk, Fox, Zapartas, de~Mink, Smith,
  Brunsden, Azalee~Bostroem, Filippenko, Shivvers, \&
  Zheng}]{ryder_ultraviolet_2018}
Ryder, S.~D., Van~Dyk, S.~D., Fox, O.~D., {et~al.} 2018, The Astrophysical
  Journal, 856, 83, \dodoi{10.3847/1538-4357/aaaf1e}

\bibitem[{{Sana} {et~al.}(2012){Sana}, {de Mink}, {de Koter}, {Langer},
  {Evans}, {Gieles}, {Gosset}, {Izzard}, {Le Bouquin}, \&
  {Schneider}}]{sana:12}
{Sana}, H., {de Mink}, S.~E., {de Koter}, A., {et~al.} 2012, Science, 337, 444,
  \dodoi{10.1126/science.1223344}

\bibitem[{{Schneider} {et~al.}(2021){Schneider}, {Podsiadlowski}, \&
  {M{\"u}ller}}]{2021A&A...645A...5S}
{Schneider}, F.~R.~N., {Podsiadlowski}, P., \& {M{\"u}ller}, B. 2021, \aap,
  645, A5, \dodoi{10.1051/0004-6361/202039219}

\bibitem[{{Shivvers} {et~al.}(2019){Shivvers}, {Filippenko}, {Silverman},
  {Zheng}, {Foley}, {Chornock}, {Barth}, {Cenko}, {Clubb}, {Fox},
  {Ganeshalingam}, {Graham}, {Kelly}, {Kleiser}, {Leonard}, {Li}, {Matheson},
  {Mauerhan}, {Modjaz}, {Serduke}, {Shields}, {Steele}, {Swift}, {Wong}, \&
  {Yuk}}]{Shivver2019}
{Shivvers}, I., {Filippenko}, A.~V., {Silverman}, J.~M., {et~al.} 2019, \mnras,
  482, 1545, \dodoi{10.1093/mnras/sty2719}

\bibitem[{{Skinner} {et~al.}(2019){Skinner}, {Dolence}, {Burrows}, {Radice}, \&
  {Vartanyan}}]{skinner2019}
{Skinner}, M.~A., {Dolence}, J.~C., {Burrows}, A., {Radice}, D., \&
  {Vartanyan}, D. 2019, \apjs, 241, 7, \dodoi{10.3847/1538-4365/ab007f}

\bibitem[{{Sravan} {et~al.}(2019){Sravan}, {Marchant}, \&
  {Kalogera}}]{Sravan+2019}
{Sravan}, N., {Marchant}, P., \& {Kalogera}, V. 2019, \apj, 885, 130

\bibitem[{{Steiner} {et~al.}(2013){Steiner}, {Hempel}, \&
  {Fischer}}]{2013ApJ...774...17S}
{Steiner}, A.~W., {Hempel}, M., \& {Fischer}, T. 2013, \apj, 774, 17,
  \dodoi{10.1088/0004-637X/774/1/17}

\bibitem[{{Sukhbold} \& {Woosley}(2014)}]{2014ApJ...783...10S}
{Sukhbold}, T., \& {Woosley}, S.~E. 2014, \apj, 783, 10,
  \dodoi{10.1088/0004-637X/783/1/10}

\bibitem[{{Sukhbold} {et~al.}(2018){Sukhbold}, {Woosley}, \&
  {Heger}}]{2018ApJ...860...93S}
{Sukhbold}, T., {Woosley}, S.~E., \& {Heger}, A. 2018, \apj, 860, 93,
  \dodoi{10.3847/1538-4357/aac2da}

\bibitem[{{Summa} {et~al.}(2016){Summa}, {Hanke}, {Janka}, {Melson}, {Marek},
  \& {M{\"u}ller}}]{summa2016}
{Summa}, A., {Hanke}, F., {Janka}, H.-T., {et~al.} 2016, \apj, 825, 6,
  \dodoi{10.3847/0004-637X/825/1/6}

\bibitem[{{Summa} {et~al.}(2018){Summa}, {Janka}, {Melson}, \&
  {Marek}}]{summa2018}
{Summa}, A., {Janka}, H.-T., {Melson}, T., \& {Marek}, A. 2018, \apj, 852, 28,
  \dodoi{10.3847/1538-4357/aa9ce8}

\bibitem[{Taddia {et~al.}(2018)Taddia, Stritzinger, Bersten, Baron, Burns,
  Contreras, Holmbo, Hsiao, Morrell, Phillips, Sollerman, \&
  Suntzeff}]{taddia_carnegie_2018}
Taddia, F., Stritzinger, M.~D., Bersten, M., {et~al.} 2018, Astronomy and
  Astrophysics, 609, A136, \dodoi{10.1051/0004-6361/201730844}

\bibitem[{{Taddia} {et~al.}(2018){Taddia}, {Stritzinger}, {Bersten}, {Baron},
  {Burns}, {Contreras}, {Holmbo}, {Hsiao}, {Morrell}, {Phillips}, {Sollerman},
  \& {Suntzeff}}]{2018A&A...609A.136T}
{Taddia}, F., {Stritzinger}, M.~D., {Bersten}, M., {et~al.} 2018, \aap, 609,
  A136, \dodoi{10.1051/0004-6361/201730844}

\bibitem[{{Tauris} {et~al.}(2015){Tauris}, {Langer}, \&
  {Podsiadlowski}}]{tauris:15}
{Tauris}, T.~M., {Langer}, N., \& {Podsiadlowski}, P. 2015, \mnras, 451, 2123,
  \dodoi{10.1093/mnras/stv990}

\bibitem[{{Tews} {et~al.}(2017){Tews}, {Lattimer}, {Ohnishi}, \&
  {Kolomeitsev}}]{2017ApJ...848..105T}
{Tews}, I., {Lattimer}, J.~M., {Ohnishi}, A., \& {Kolomeitsev}, E.~E. 2017,
  \apj, 848, 105, \dodoi{10.3847/1538-4357/aa8db9}

\bibitem[{{The LIGO Scientific Collaboration} {et~al.}(2020){The LIGO
  Scientific Collaboration}, {the Virgo Collaboration}, {Abbott}, {Abbott},
  {Abraham}, {Acernese}, {Ackley}, {Adams}, {Adhikari}, {Adya}, {Affeldt},
  {Agathos}, {Agatsuma}, {Aggarwal}, {Aguiar}, {Aich}, {Aiello}, {Ain},
  {Ajith}, {Akcay}, {Allen}, {Allocca}, \& et~al.}]{2020arXiv200612611T}
{The LIGO Scientific Collaboration}, {the Virgo Collaboration}, {Abbott}, R.,
  {et~al.} 2020, arXiv e-prints, arXiv:2006.12611.
\newblock \doarXiv{2006.12611}

\bibitem[{{Timmes}(1999)}]{timmes1999}
{Timmes}, F.~X. 1999, \apjs, 124, 241, \dodoi{10.1086/313257}

\bibitem[{{Timmes} {et~al.}(2000){Timmes}, {Hoffman}, \&
  {Woosley}}]{timmes2000}
{Timmes}, F.~X., {Hoffman}, R.~D., \& {Woosley}, S.~E. 2000, \apjs, 129, 377,
  \dodoi{10.1086/313407}

\bibitem[{{Utrobin} {et~al.}(2021){Utrobin}, {Wongwathanarat}, {Janka},
  {Mueller}, {Ertl}, {Menon}, \& {Heger}}]{2021arXiv210209686U}
{Utrobin}, V.~P., {Wongwathanarat}, A., {Janka}, H.~T., {et~al.} 2021, arXiv
  e-prints, arXiv:2102.09686.
\newblock \doarXiv{2102.09686}

\bibitem[{{Vartanyan} \& {Burrows}(2020)}]{2020ApJ...901..108V}
{Vartanyan}, D., \& {Burrows}, A. 2020, \apj, 901, 108,
  \dodoi{10.3847/1538-4357/abafac}

\bibitem[{{Vartanyan} {et~al.}(2019{\natexlab{a}}){Vartanyan}, {Burrows}, \&
  {Radice}}]{2019MNRAS.489.2227V}
{Vartanyan}, D., {Burrows}, A., \& {Radice}, D. 2019{\natexlab{a}}, \mnras,
  489, 2227, \dodoi{10.1093/mnras/stz2307}

\bibitem[{{Vartanyan} {et~al.}(2018){Vartanyan}, {Burrows}, {Radice},
  {Skinner}, \& {Dolence}}]{vartanyan2018a}
{Vartanyan}, D., {Burrows}, A., {Radice}, D., {Skinner}, M.~A., \& {Dolence},
  J. 2018, \mnras, 477, 3091, \dodoi{10.1093/mnras/sty809}

\bibitem[{{Vartanyan} {et~al.}(2019{\natexlab{b}}){Vartanyan}, {Burrows},
  {Radice}, {Skinner}, \& {Dolence}}]{vartanyan2018b}
---. 2019{\natexlab{b}}, \mnras, 482, 351, \dodoi{10.1093/mnras/sty2585}

\bibitem[{{Vaytet} {et~al.}(2011){Vaytet}, {Audit}, {Dubroca}, \&
  {Delahaye}}]{2011JQSRT.112.1323V}
{Vaytet}, N.~M.~H., {Audit}, E., {Dubroca}, B., \& {Delahaye}, F. 2011, \jqsrt,
  112, 1323, \dodoi{10.1016/j.jqsrt.2011.01.027}

\bibitem[{{Woosley}(2019)}]{woosley2019}
{Woosley}, S.~E. 2019, \apj, 878, 49, \dodoi{10.3847/1538-4357/ab1b41}

\bibitem[{{Woosley} {et~al.}(1993){Woosley}, {Langer}, \&
  {Weaver}}]{1993ApJ...411..823W}
{Woosley}, S.~E., {Langer}, N., \& {Weaver}, T.~A. 1993, \apj, 411, 823,
  \dodoi{10.1086/172886}

\bibitem[{Woosley {et~al.}(2020)Woosley, Sukhbold, \& Janka}]{Woosley_2020}
Woosley, S.~E., Sukhbold, T., \& Janka, H.-T. 2020, The Astrophysical Journal,
  896, 56, \dodoi{10.3847/1538-4357/ab8cc1}

\bibitem[{{Yoon} {et~al.}(2017){Yoon}, {Dessart}, \&
  {Clocchiatti}}]{2017ApJ...840...10Y}
{Yoon}, S.-C., {Dessart}, L., \& {Clocchiatti}, A. 2017, \apj, 840, 10,
  \dodoi{10.3847/1538-4357/aa6afe}

\bibitem[{{Yoon} {et~al.}(2010){Yoon}, {Woosley}, \&
  {Langer}}]{2010ApJ...725..940Y}
{Yoon}, S.~C., {Woosley}, S.~E., \& {Langer}, N. 2010, \apj, 725, 940,
  \dodoi{10.1088/0004-637X/725/1/940}

\bibitem[{{Zapartas} {et~al.}(2021){Zapartas}, {de Mink}, {Justham}, {Smith},
  {Renzo}, \& {de Koter}}]{2021A&A...645A...6Z}
{Zapartas}, E., {de Mink}, S.~E., {Justham}, S., {et~al.} 2021, \aap, 645, A6,
  \dodoi{10.1051/0004-6361/202037744}

\bibitem[{{Zapartas} {et~al.}(2017){Zapartas}, {de Mink}, {Van Dyk}, {Fox},
  {Smith}, {Bostroem}, {de Koter}, {Filippenko}, {Izzard}, {Kelly}, {Neijssel},
  {Renzo}, \& {Ryder}}]{2017ApJ...842..125Z}
{Zapartas}, E., {de Mink}, S.~E., {Van Dyk}, S.~D., {et~al.} 2017, \apj, 842,
  125, \dodoi{10.3847/1538-4357/aa7467}

\bibitem[{Zapartas {et~al.}(2019)Zapartas, de~Mink, Justham, Smith, de~Koter,
  Renzo, Arcavi, Farmer, Götberg, \& Toonen}]{zapartas_diverse_2019}
Zapartas, E., de~Mink, S.~E., Justham, S., {et~al.} 2019, Astronomy and
  Astrophysics, 631, A5, \dodoi{10.1051/0004-6361/201935854}

\end{thebibliography}

\begin{figure*}[!htbp]%
\centering
     \includegraphics[width=0.8\textheight]{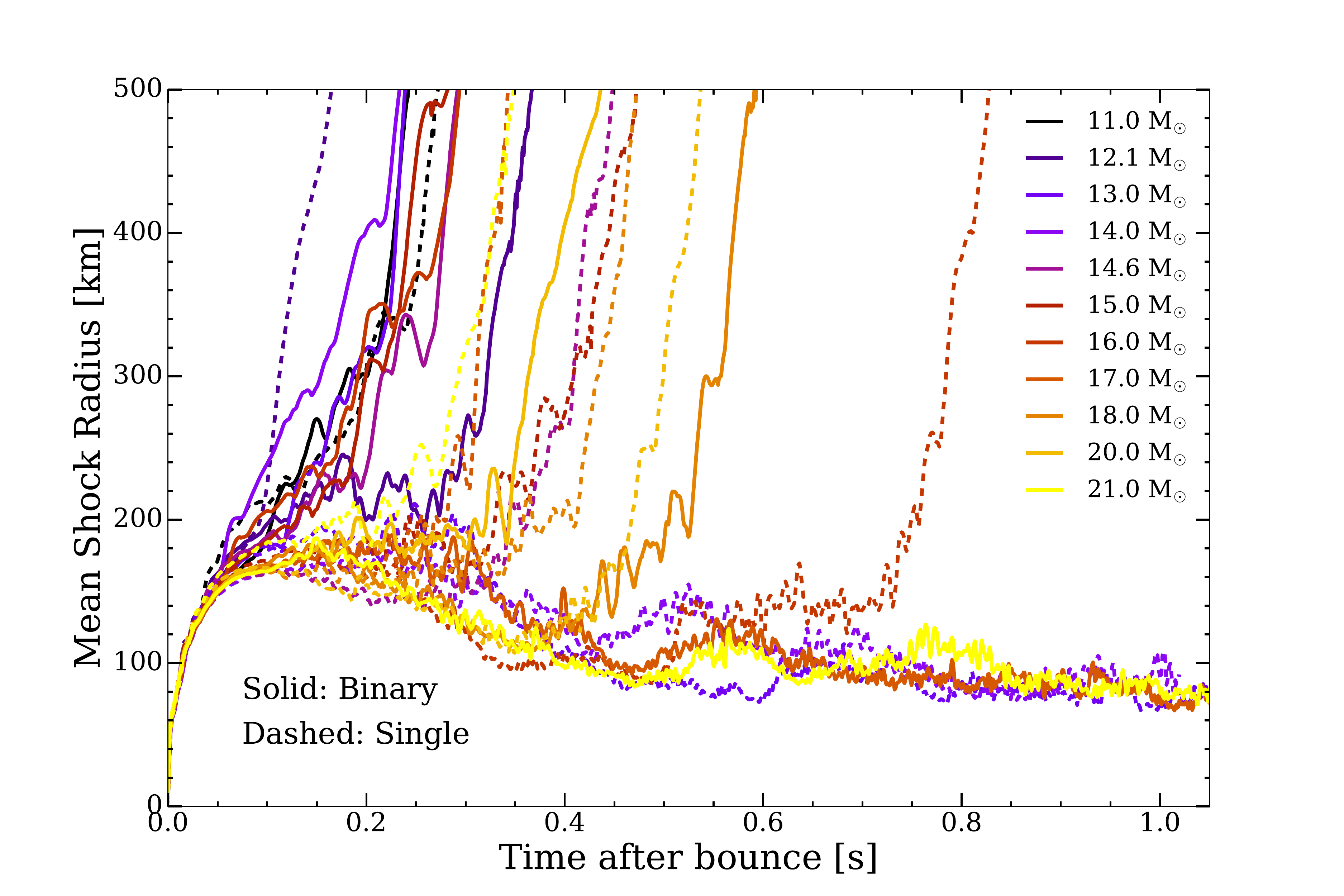}
\caption{Mean shock radii (km) for the 11 pairs of models studied as a function of time after bounce (s). The single-star models are illustrated with solid lines, and the binary-stripped models with dashed lines. All but 4 models explode, and the explosions occur within 100-800 ms post-bounce. Note that even the non-exploding models feature a bump in the shock radii after 500 ms, corresponding to the accretion of the Si/O interface. }
\label{fig:shock}
\end{figure*}

\begin{figure*}[!htbp]%
    \centering
     \includegraphics[width=0.3\textwidth]{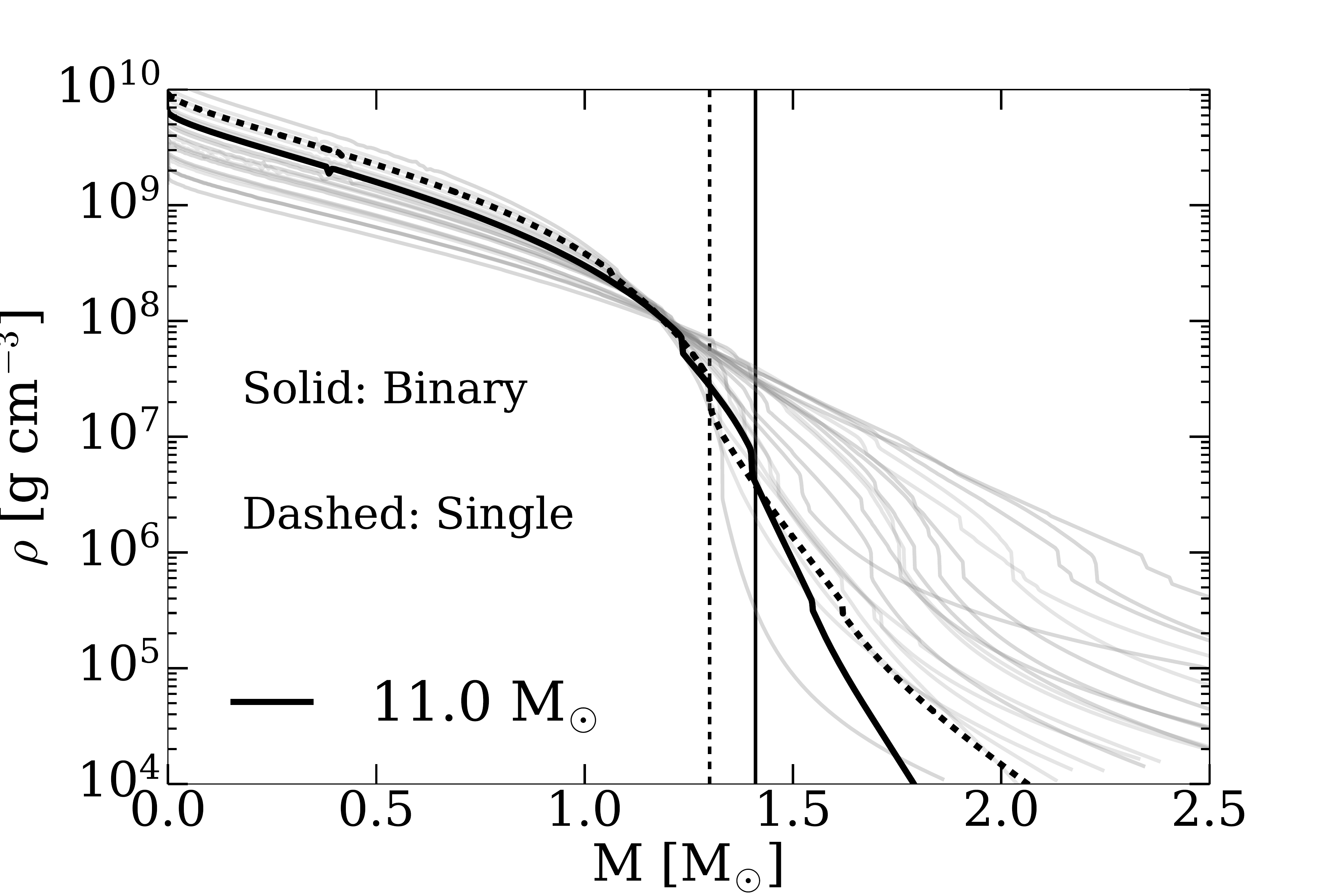}
    \includegraphics[width=0.3\textwidth]{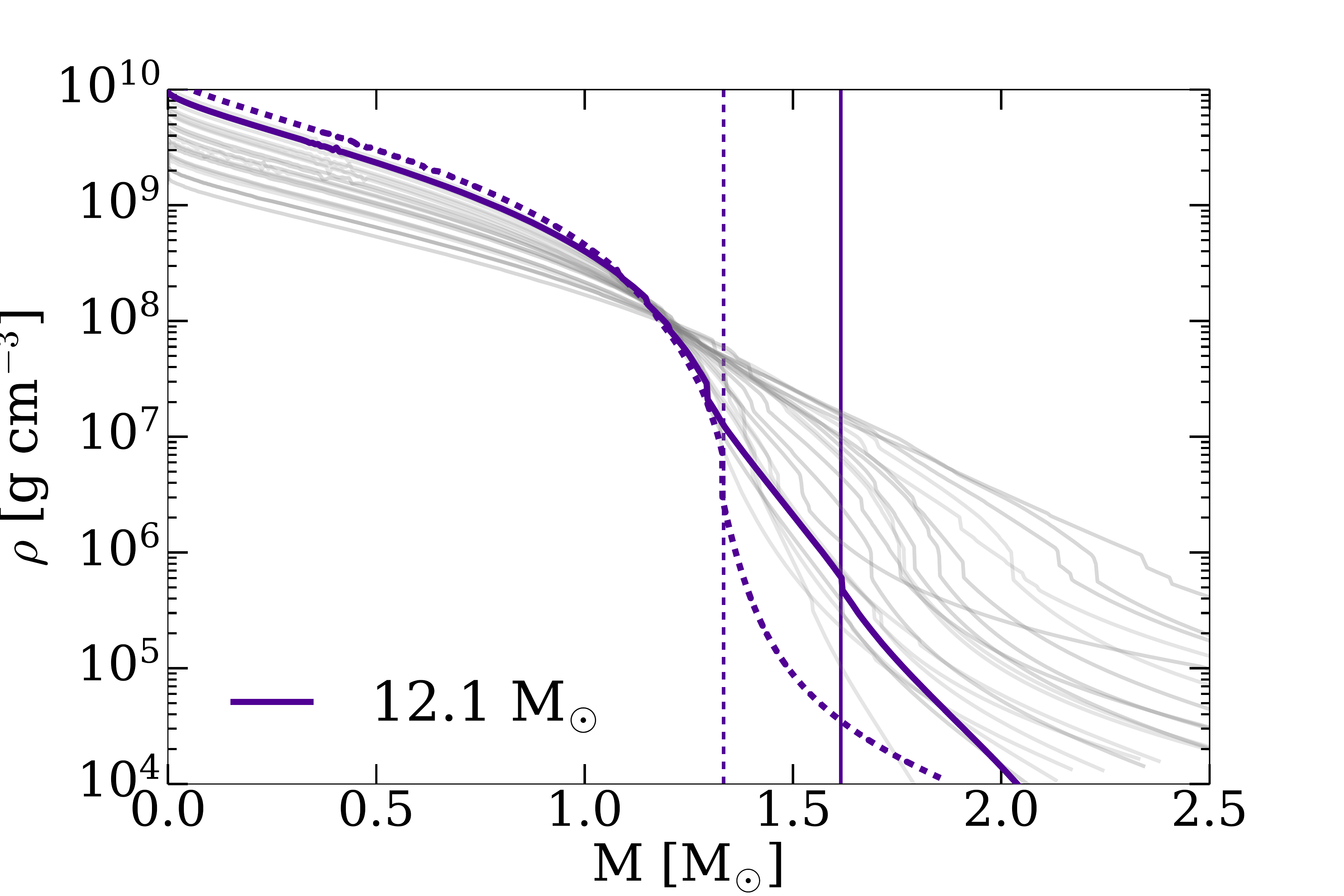}   
     \includegraphics[width=0.3\textwidth]{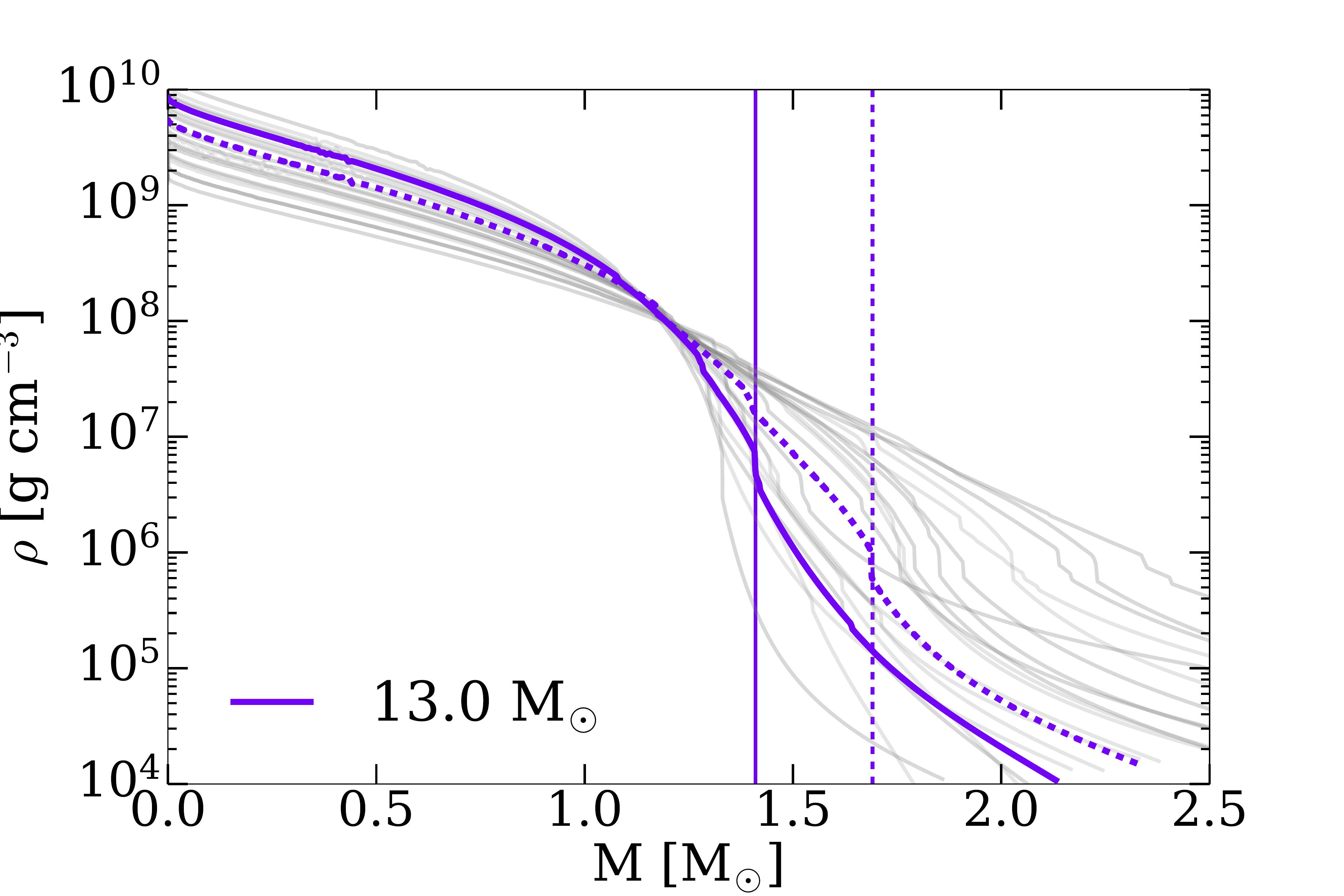}
     \vspace{0.1cm}
    \includegraphics[width=0.3\textwidth]{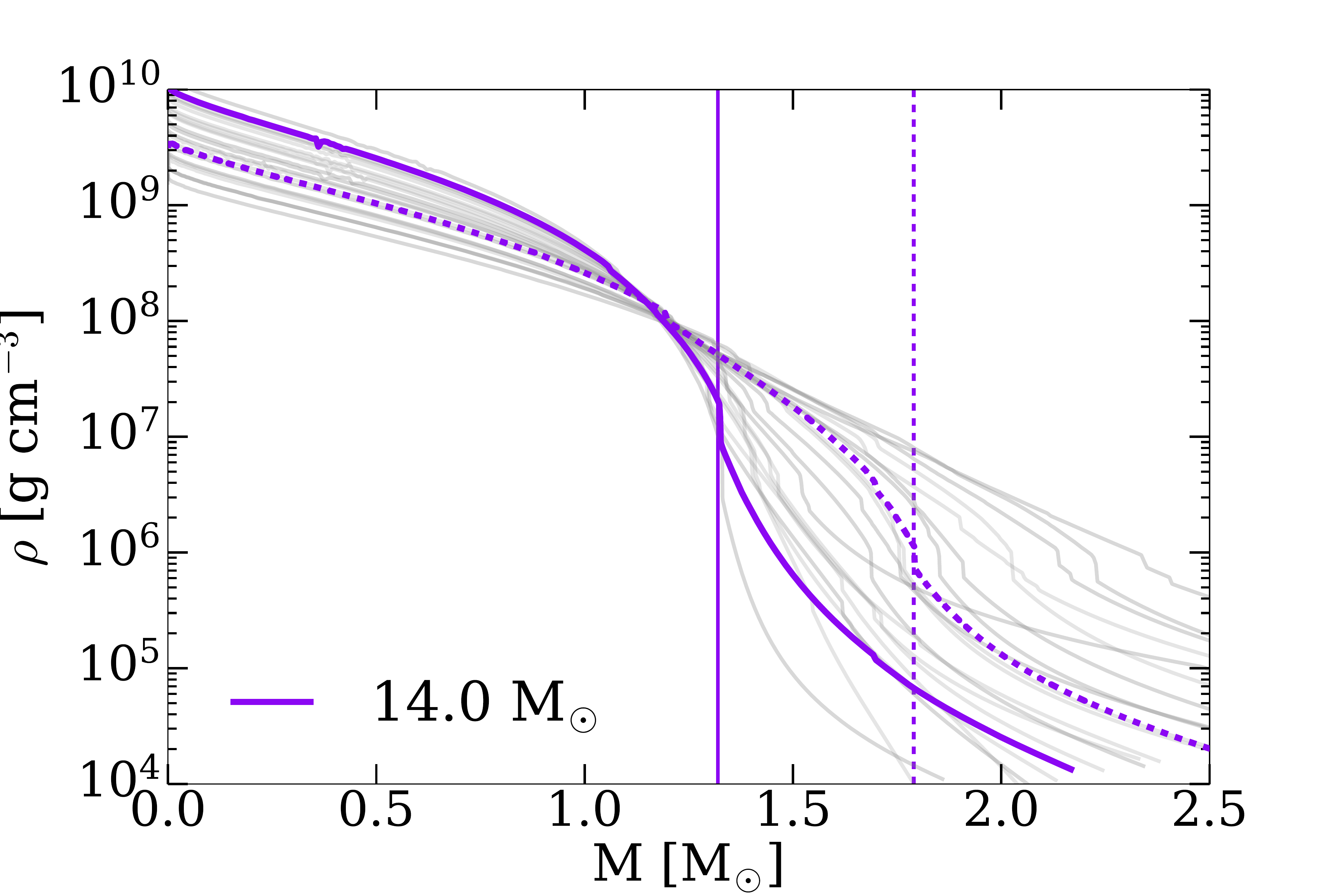}   
    \includegraphics[width=0.3\textwidth]{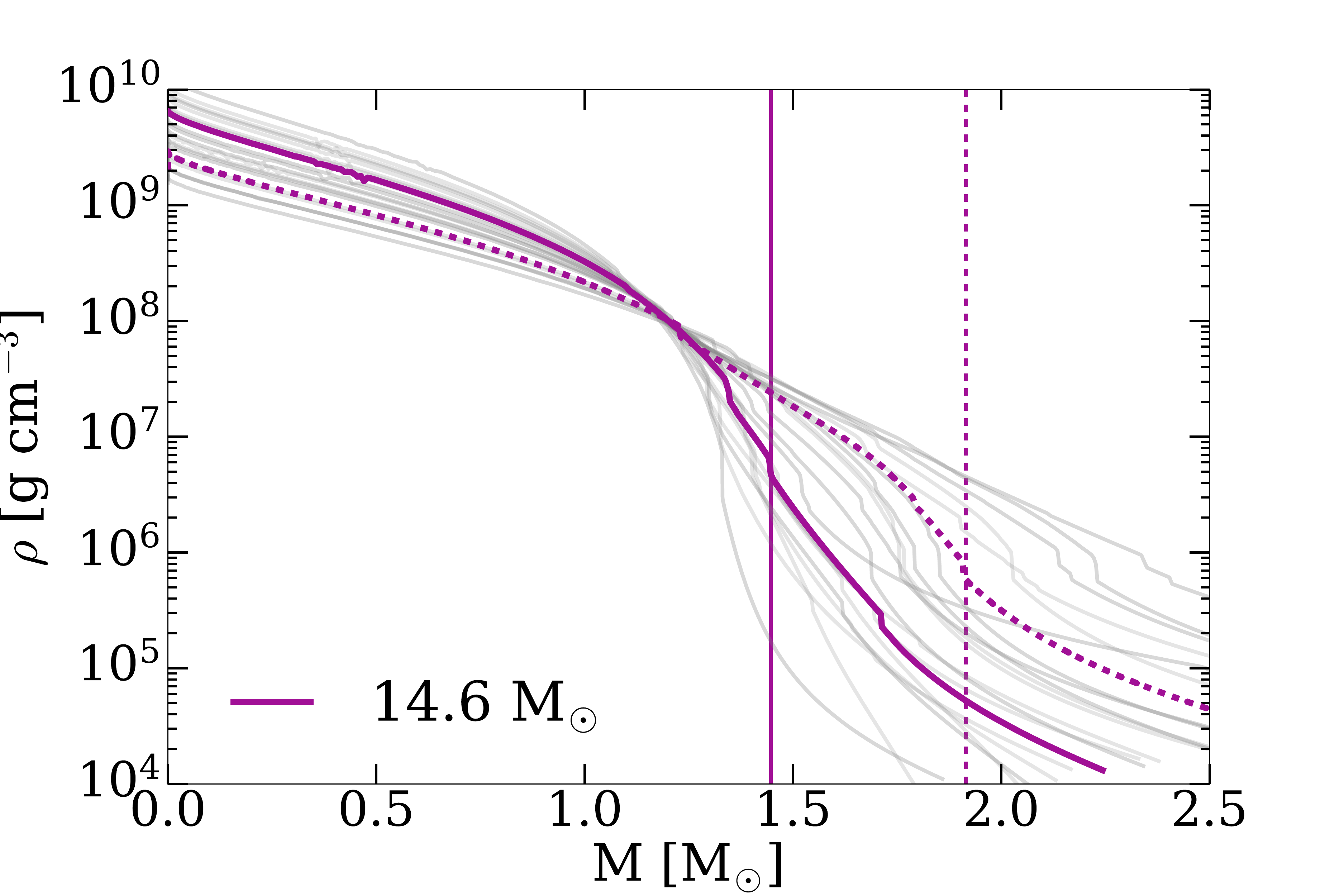}   
        \includegraphics[width=0.3\textwidth]{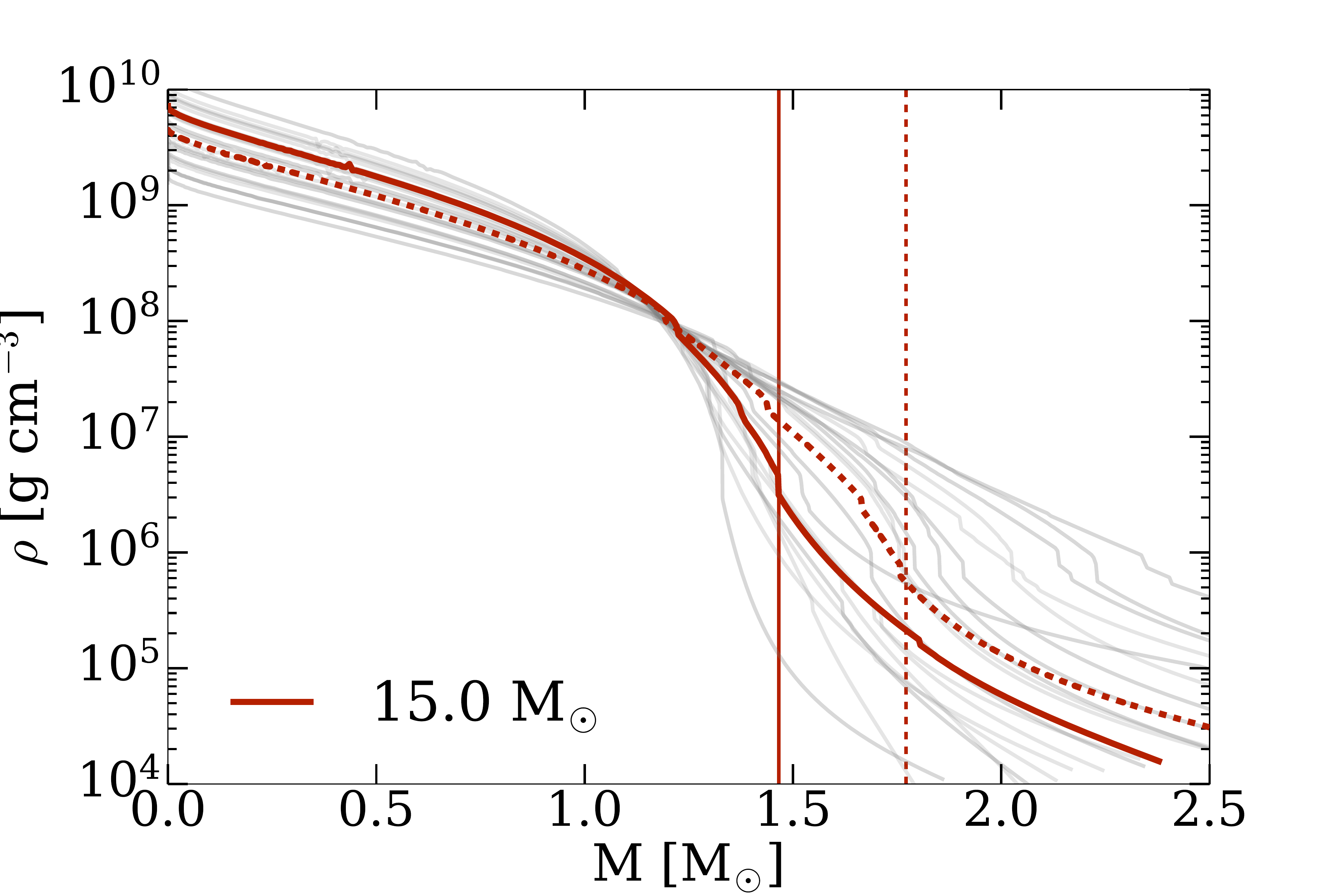}   
     \vspace{0.1cm}            \includegraphics[width=0.3\textwidth]{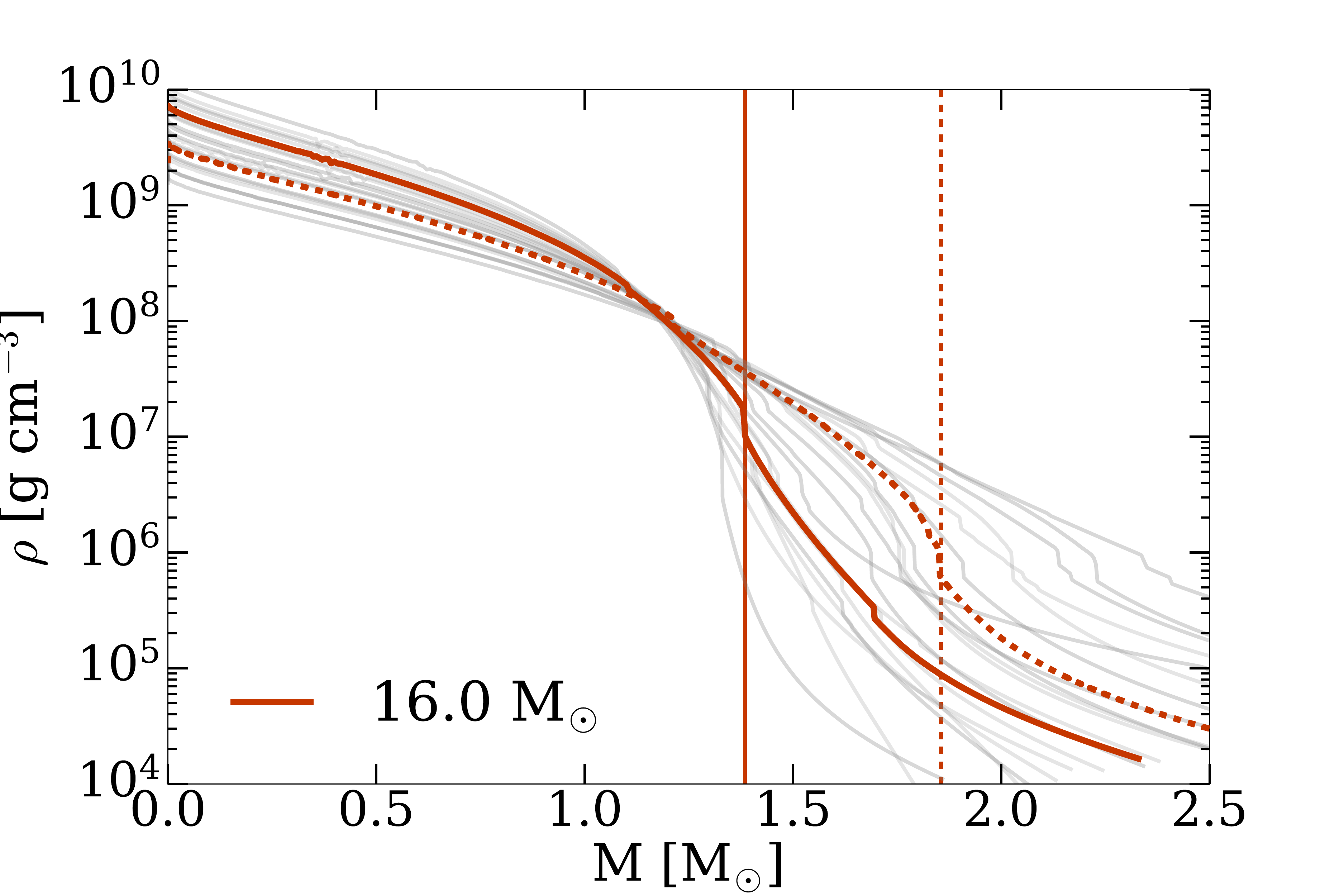}  
    \includegraphics[width=0.3\textwidth]{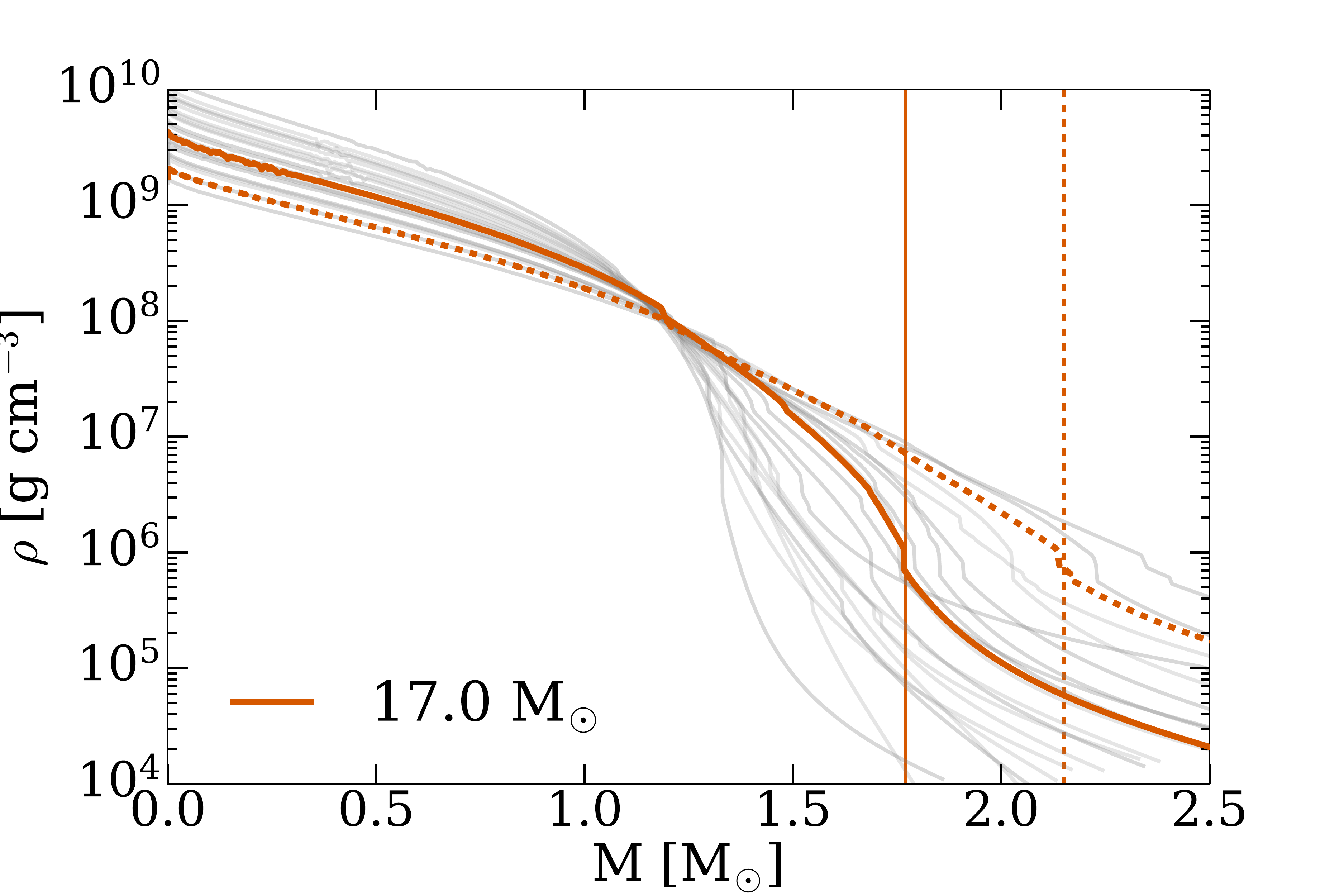}   
    \includegraphics[width=0.31\textwidth]{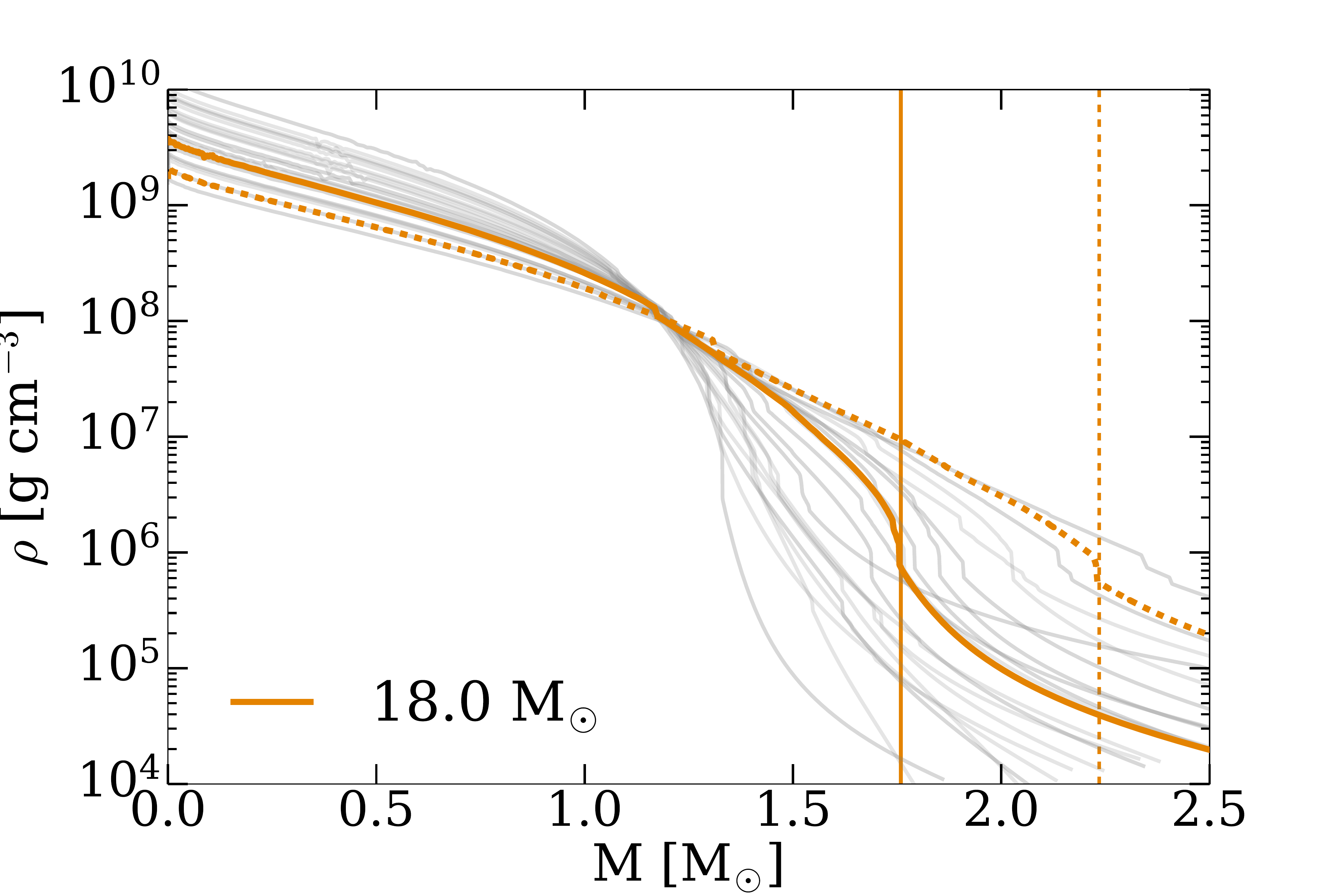}    
         \vspace{0.1cm}    \includegraphics[width=0.31\textwidth]{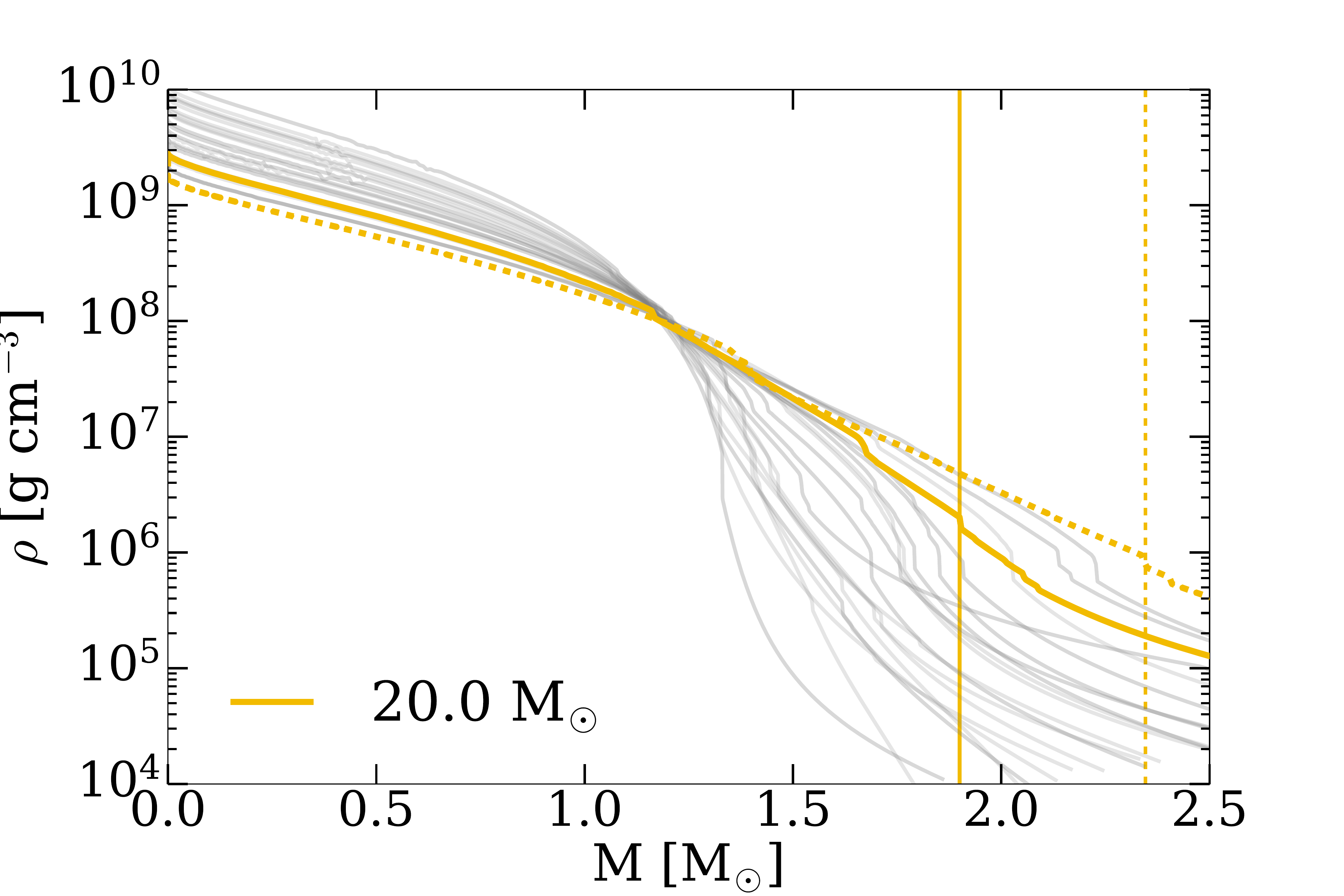}   
            \includegraphics[width=0.31\textwidth]{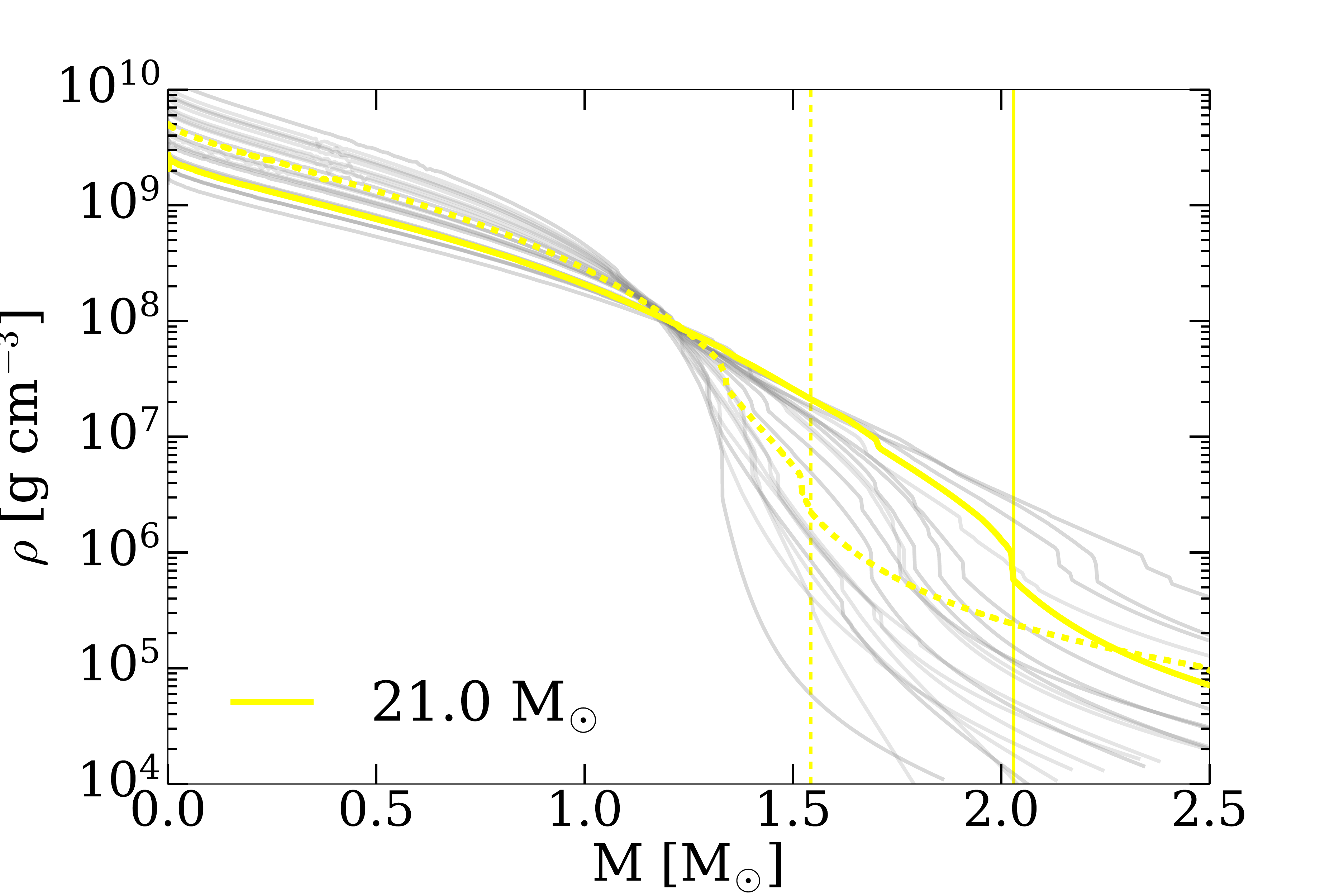}

	\caption{Density profiles for each of the 11 pairs of progenitors studied here. Dashed lines indicate single star progenitors and solid lines indicate binary-stripped. We overplot as vertical lines where the relevant density drops, indicative of a sharp Si/O interface, occur. Note that several models have multiple sharp interfaces $-$ however, the interior Fe/Si interface is often irrelevant to explosion because it is accreted onto the PNS during core-collapse and lies interior to the post-bounce shock. In other cases where the relevant interface may be fragmented, we highlight the interface whose accretion occurs on timescales relevant to shock revival. Note the presence and strength of an interface is further complicated by limitations of mixing-length theory, particularly when nuclear burning occurs on similar timescales to convective turnover.}
	\label{fig:rho_M_rshock}
\end{figure*}

\appendix\label{app}

In the text, we highlighted the 14-M$_{\odot}$ progenitor in the text as a case study of binary-stripped vs single star profile and explosion outcome, and discussed the various exceptions. Here, we illustrate the progenitor profile and explosion outcome of all models. In Fig.\,\ref{fig:shock}, we plot the shock radii as function of time for all the models studied here. Only four models show no explosion, with all the other moodels evincing explosion within the first second post-bounce. Shock revival corresponds to   the presence of a sharp compositional interface located sufficiently deep within the star.

In Fig.\,\ref{fig:rho_M_rshock}, we plot the density profiles for each of 11 pairs of single and binary-stripped models studied here. We highlight the location of the Si/O or equivalent density interface for each model as a vertical line. To identify the location of the relevant compositional interface, we look for a sharp density drop of a factor of several around 1.5-2 M$_{\odot}$, at a density of a few million g cm$^{-3}$. We then correlate these with the composition (e.g. top left panel of Fig.\,\ref{fig:typical}) to isolate which compositional boundary the interface corresponds to. When there are several ``fragmented" interfaces in close proximity, we then check that the accretion of the chosen interface coincides with shock revival, or in the case of a failed explosion, a bump in the shock radii.

\clearpage

\end{document}